\documentclass[10pt]{article}
\usepackage{typearea}\typearea{14}
\usepackage{epsfig,amsmath,amsfonts,amssymb,upgreek,dsfont}
\newtheorem{theorem}{Theorem}[section]
\newtheorem{lemma}[theorem]{Lemma}
\newtheorem{corollary}[theorem]{Corollary}

\newtheorem{definition}[theorem]{Definition}

\makeatletter
\@addtoreset{equation}{section}
\makeatother

\makeatletter
\long\def\@makecaption#1#2{{\small
\advance\leftskip1cm
\advance\rightskip1cm
\vskip\abovecaptionskip
\sbox\@tempboxa{#1: #2}%
\ifdim \wd\@tempboxa >\hsize
 #1: #2\par
\else
\global \@minipagefalse
\hb@xt@\hsize{\hfil\box\@tempboxa\hfil}%
\fi
\vskip\belowcaptionskip}}
\makeatother
\def\eq#1\en{\begin{equation}#1\end{equation}}  
\def\eqa#1\ena{\begin{align}#1\end{align}}
\def\eqg#1\eng{\begin{gather}#1\end{gather}}
\newcommand{\lb}[1]{\label{e:#1}}
\newcommand{\rlb}[1]{\eqref{e:#1}} 
\newcommand{\nl}{\notag\\}

\newcommand{\norms}[1]{\Vert#1\Vert}

\newcommand{\bkt}[1]{\left\langle#1\right\rangle}

\newcommand{\sumtwo}[2]%
{\mathop{\sum_{#1}}_{#2}}
\newcommand{\sumthree}[3]%
{\mathop{\mathop{\sum_{#1}}_{#2}}_{#3}}
\newcommand{\sumfour}[4]%
{\mathop{\mathop{\mathop{\sum_{#1}}_{#2}}_{#3}}_{#4}} 
\newcommand{\prodtwo}[2]%
{\mathop{\prod_{#1}}_{#2}}
\newcommand{\mintwo}[2]%
{\mathop{\min_{#1}}_{#2}}
\newcommand{\maxtwo}[2]%
{\mathop{\max_{#1}}_{#2}}
\newcommand{\maxthree}[3]%
{\mathop{\mathop{\max_{#1}}_{#2}}_{#3}}
\newcommand{\limtwo}[2]%
{\mathop{\lim_{#1}}_{#2}}
\newcommand{\suptwo}[2]%
{\mathop{\sup_{#1}}_{#2}}
\newcommand{\supthree}[3]%
{\mathop{\mathop{\sup_{#1}}_{#2}}_{#3}}
\newcommand{\supfour}[4]%
{\mathop{\mathop{\mathop{\sup_{#1}}_{#2}}_{#3}}_{#4}} 
\newcommand{\inftwo}[2]%
{\mathop{\inf_{#1}}_{#2}}
\newcommand{\infthree}[3]%
{\mathop{\mathop{\inf_{#1}}_{#2}}_{#3}}
\newcommand{\inffour}[4]%
{\mathop{\mathop{\mathop{\inf_{#1}}_{#2}}_{#3}}_{#4}} 
\newcommand\calH{{\cal H}}

\newcommand\calT{{\cal T}}

\newcommand{\mrx}{\mathrm{x}}
\newcommand{\mry}{\mathrm{y}}
\newcommand{\mrz}{\mathrm{z}}
\newcommand{\bbC}{\mathbb{C}}

\newcommand{\bbN}{\mathbb{N}}

\newcommand{\bbR}{\mathbb{R}}
\newcommand{\bbZ}{\mathbb{Z}}
\newcommand{\up}{\uparrow}

\newcommand{\Di}{\mathit{\Delta}}
\newcommand{\qedm}{\rule{1.5mm}{3mm}}
\newcommand{\hS}{\hat{S}}
\newcommand{\hA}{\hat{A}}
\newcommand{\hB}{\hat{B}}
\newcommand{\hH}{\hat{H}}
\newcommand{\hQ}{\hat{Q}}
\newcommand{\hU}{\hat{U}}
\newcommand{\hV}{\hat{V}}
\newcommand{\hW}{\hat{W}}
\newcommand{\hh}{\hat{h}}
\newcommand{\hu}{\hat{u}}

\newcommand{\hSa}{\hat{S}^{(\alpha)}}
\newcommand{\hSx}{\hat{S}^{(\mrx)}}
\newcommand{\hSy}{\hat{S}^{(\mry)}}
\newcommand{\hSz}{\hat{S}^{(\mrz)}}

\newcommand{\hUx}{\hU_{x,\ell}}

\newcommand{\hilb}{\mathfrak{h}}
\newcommand{\bra}[1]{\langle#1|}
\newcommand{\ket}[1]{|#1\rangle}

\newcommand{\ZZ}{\bbZ_2\times\bbZ_2}

\newcommand{\OA}{\mathfrak{A}}
\newcommand{\Aloc}{\mathfrak{A}_{\rm loc}}
\newcommand{\iop}{\hat{1}}
\newcommand{\DE}{\Di E}
\newcommand{\LaL}{\Lambda_L}
\newcommand{\CL}{\mathcal{C}_L}
\newcommand{\GS}{\ket{\Phi_{\rm GS}}}
\newcommand{\GSb}{\bra{\Phi_{\rm GS}}}
\newcommand{\snorm}{\norms}

\newcommand{\pjj}{(j_1,j_2)}
\newcommand{\jj}{j_1,j_2}

\newcommand{\Uo}{\mathrm{U(1)}}

\newcommand{\ind}{\operatorname{ind}}
\newcommand{\Ind}{\operatorname{Ind}}

\newcommand{\hux}{\hu^{(\mrx)}}
\newcommand{\huy}{\hu^{(\mry)}}
\newcommand{\huz}{\hu^{(\mrz)}}
\newcommand{\hug}{\hu^{(g)}}
\newcommand{\huh}{\hu^{(h)}}

\newcommand{\xyz}{\mrx,\mry,\mrz}
\newcommand{\oL}{\{1,\ldots,L\}}

\newcommand{\id}{\mathds{1}}

\newcommand{\Hw}{H_{\rm GNS}}
\newcommand{\Hiw}{\mathcal{H}}
\newcommand{\piw}{\uppi}
\newcommand{\Ow}{\Omega}
\newcommand{\bktw}[1]{\langle#1\rangle}

\newcommand{\re}{\mathrm{e}}

\begin{document}

%\begin{flushright}
%\footnotesize
%Draft version 2 (not yet for publication), February 3, 2022.
%\end{flushright}
%\vfil 

\noindent
{\Large\bf The Lieb-Schultz-Mattis Theorem}

\vspace{1mm}
\renewcommand{\thefootnote}{\fnsymbol{footnote}}

\noindent
{\large\bf A Topological Point of View}\footnote{%
Published in Rupert L. Frank, Ari Laptev, Mathieu Lewin, and Robert Seiringer eds. ``The Physics and Mathematics of Elliott Lieb'' vol.~2, pp.~405--446 (European Mathematical Society Press, 2022).
}

\medskip\noindent
Hal Tasaki\footnote{%
Department of Physics, Gakushuin University, Mejiro, Toshima-ku, 
Tokyo 171-8588, Japan.
}
\renewcommand{\thefootnote}{\arabic{footnote}}
\setcounter{footnote}{0}

\begin{quotation}
\small\noindent
We review the Lieb-Schultz-Mattis theorem and its variants, which are no-go theorems that state that a quantum many-body system with certain conditions cannot have a locally-unique gapped ground state.
We restrict ourselves to one-dimensional quantum spin systems and discuss both the generalized Lieb-Schultz-Mattis theorem for models with U(1) symmetry and the extended Lieb-Schultz-Mattis theorem for models with discrete symmetry.
We also discuss the implication of the same arguments to systems on the infinite cylinder, both with the periodic boundary conditions and with the spiral boundary conditions.

For models with U(1) symmetry, we here present a rearranged version of the original proof of Lieb, Schultz, and Mattis based on the twist operator.
As the title suggests we take a modern topological point of view and prove the generalized Lieb-Schultz-Mattis theorem by making use of a topological index (which coincides with the filling factor).
By a topological index, we mean an index that characterizes a locally-unique gapped ground state and is invariant under continuous (or smooth) modification of the ground state.

For models with discrete symmetry, we describe the basic idea of the most general proof based on the topological index introduced in the context of symmetry-protected topological phases.
We start from background materials such as the classification of projective representations of the symmetry group.

We also review the notion that we call a locally-unique gapped ground state of a quantum spin system on an infinite lattice and present basic theorems.
This notion turns out to be natural and useful from the physicists' point of view.

We have tried to make the present article readable and almost self-contained.
We only assume basic knowledge about quantum spin systems.

\smallskip

\em\small\noindent
Dedicated to Elliott Lieb on the occasion of his 90th birthday.
\end{quotation}

\vfil

\tableofcontents

\vfil

\newpage
%%%%%%%%%%%%%%%%%%
\section{Introduction}
\label{s:intro}
The Lieb-Schultz-Mattis theorem first appeared as Theorem~2 in Appendix~B of \cite{LSM}.
The main topic of this paper was exactly solvable spin chains, and the theorem was stated as a remark about the nature of the ground state and low-energy excited states.
The theorem was proved by a simple variational argument that makes use of the unitary transformation described by the twist operator.
The twist operator plays an essential role also in the present article.

These days, the term Lieb-Schultz-Mattis theorem, often abbreviated as the LSM theorem, stands for a general no-go theorem that states that a certain quantum many-body system cannot have a unique gapped ground state.\footnote{
The system may have multiple ground states or a ground state accompanied by gapless excitations.
Being a no-go theorem, an LSM-type theorem, in general, does not deal with the precise nature of the ground state(s).
}
Such statements are important in the study of topological phases of quantum matter, where unique gapped ground states are the main objects of study.
See, e.g., \cite{ZengChenZhouWenBOOK} and Part~II of \cite{TasakiBook}.

In the present article\footnote{
This article is basically a review, but some results have not been discussed in the literature.
}, we restrict ourselves to one-dimensional translation-invariant quantum spin systems and discuss a full generalization of the original LSM theorem for models with continuous U(1) symmetry and the extended LSM theorem for models with discrete $\ZZ$ symmetry.
We shall treat systems on the infinitely long cylinder as well as on the infinite chain.
As the title suggests, we take a viewpoint that is inspired by recent ``topological condensed matter physics'' and present proofs based on topological indices.
We also discuss in detail the important and useful notion of a locally-unique gapped ground state and present basic theorems.

Although we have tried to make the present article almost self-contained, we do assume basic knowledge about quantum spin systems.
We recommend Chapter~2 of \cite{TasakiBook} for a compact introduction.

\paragraph*{Basic strategies}  
Before going into details, we briefly discuss general strategies of the proofs of LSM-type theorems.

In the original proof of the theorem for the $S=1/2$ antiferromagnetic Heisenberg model \rlb{HAF} on the periodic chain with $L$ sites, where $L$ is even, one starts by noting that the model has a unique ground state $\GS$.
The uniqueness, along with the rotation invariance of the Hamiltonian, readily implies that $\exp[-i\theta\sum_{j=1}^L\hSz_j]\GS=\GS$ for any $\theta$, i.e., the ground state is invariant under any uniform rotation of spins about the z-axis.
(See section~\ref{s:setting1D} for the notation.)
One then defines a trial state $\ket{\Psi}=\hU_{\rm twist}\GS$, where
\eq
\hU_{\rm twist}=\exp\Bigl[-i\sum_{j=1}^L\frac{2\pi j}{L}\,\hSz_j\Bigr],
\lb{Utwist}
\en
is the twist operator.
It rotates the spin at $j$ by angle $\theta_j=2\pi j/L$ about the z-axis.\footnote{We recall that Bloch introduced the same twist operator to study superconductivity \cite{Bohm,TadaKoma,Watanabe2019}.}
Since the rotation angle varies only gradually, the twist operator $\hU_{\rm twist}$ leaves the ground state almost unchanged, at least locally.
This observation leads to a simple variational estimate
\eq
\bra{\Psi}\hH_{\rm HAF}\ket{\Psi}-E_{\rm GS}
=\bra{\Phi_{\rm GS}}(\hU_{\rm twist}^\dagger\hH_{\rm HAF}\hU_{\rm twist}-\hH_{\rm HAF})\GS
\le\frac{\rm const.}{L},
\lb{PsiLSM}
\en
where $\hH_{\rm HAF}$ is the Hamiltonian and $E_{\rm GS}$ is the ground state energy.
See the proof of Lemma~\ref{l:var}.
We see that the energy of the trial state $\ket{\Psi}$ is only slightly higher than the ground state energy.
This estimate, along with the orthogonality $\bkt{\Phi_{\rm GS}|\Psi}=0$ (which of course should be proved, and is indeed essential), shows that the energy gap immediately above the ground state energy does not exceed ${\rm const.}/L$.
Noting that $L$ can be made as large as one wishes, we find that there is no energy gap immediately above the ground state energy.

In this manner, the no-go theorem for a unique gapped ground state is proved by explicit construction of low-energy excited states.

In his paper that initiated the LSM-type theorems for higher-dimensional models, Oshikawa \cite{O} showed that LSM-type statements may be obtained directly by examining the properties of gapped ground states.
This leads to a different strategy for proving LSM-type no-go theorems in which one assumes that the model has a unique gapped ground state and derives some characteristic property of the model.
The property is a necessary condition for the existence of a unique gapped ground state.
Then the contraposition gives an LSM-type no-go theorem, i.e., the model cannot have a unique gapped ground state if the above condition is not satisfied.

Comparing the two strategies, the first variational strategy is more informative since one gets knowledge about low-energy excitations.
If one is only interested in proving a no-go theorem, on the other hand, the second strategy is sufficient and, in many cases, more efficient.
The recent extended LSM theorem for models with discrete symmetry (which we review in section~\ref{s:ZZ}) is so far proved only by employing the second strategy.

\paragraph*{The approach in the present article}  
In the present article, we discuss both the generalization of the original LSM theorem for models with U(1) symmetry (section~\ref{s:U1}) and the extended LSM theorem for models with discrete $\ZZ$ symmetry (section~\ref{s:ZZ}) by employing the second strategy discussed above and also by making use of topological indices.
By a topological index, we mean an index that is associated with a locally-unique gapped ground state and is invariant under continuous (or smooth) modification of the ground state.
The index that we use for models with discrete symmetry is the one introduced to classify symmetry-protected topological phases.
For both the theorems, we focus on one-dimensional models that have translation invariance\footnote{
One can also prove LSM-type theorems for models with reflection invariance.
See \cite{Fuji,PWJZ} and Theorem~4 and Appendix~B of \cite{OTT}.
}, and also discuss results for models on the infinite cylinder (section~\ref{s:2D}).

For models with U(1) symmetry, this approach based on the second strategy (where an explicit variational estimate is avoided) is not standard, but we found that it simplifies the proof of the most general version of the LSM theorem.
As one application of their index theorem, Bachmann, Bols, De Roeck, and Fraas \cite{BachmannBolsDeRoecFraas2019} gave a new proof of the generalized LSM theorem for U(1) symmetric models by using a strategy similar to ours.
But we note that their new argument is much stronger than ours,  which is after all a rearrangement of the traditional proof that goes back to Lieb, Schultz, and Mattis \cite{LSM}.
In particular, the theorem of Bachmann, Bols, De Roeck, and Fraas  \cite{BachmannBolsDeRoecFraas2019}  applies also to higher-dimensional models.
See section~\ref{s:discussion}.

We here point out that both the generalized LSM theorem and the extended LSM theorem do not only rule out a unique gapped ground state but also rule out a slightly more general class of ground states that we call locally-unique gapped ground states.
This extension is practically important since the infinite volume limit of a sequence of unique gapped ground states of finite systems is not necessarily a unique gapped ground state but is always a locally-unique gapped ground state.
We can say that a locally-unique gapped ground state is a natural notion from the physicists' point of view.
The extension may be also conceptually interesting since a locally-unique gapped ground state may exhibit spontaneous symmetry breaking.
We shall discuss the notion of a locally-unique gapped ground state in detail in section~\ref{s:Gen} and present Ogata's proof of the basic theorem in Appendix~\ref{s:Yoshiko}.

%%%%%%%%%%%%%%%

\section{General setting}
\label{s:Gen}

Throughout the present article, we consider quantum spin systems on the infinite chain $\bbZ$ or the infinitely long cylinder $\LaL=\bbZ\times\oL$.
The treatment of infinite systems is not standard in the physics literature, but it has a clear advantage in discussing LSM-type theorems.
Recall that, in a finite system, a unique ground state is necessarily accompanied by a nonzero energy gap.
One needs to examine the size dependence of the gap to determine whether the ground state (in the infinite volume limit) has a nonzero gap.\footnote{%
\label{fn:finite}
{\em Note to experts:}\/ There are further, probably more essential, problems in discussing LSM-type theorems in finite systems.

First, it may happen that distinct (i.e., mutually orthogonal) ground states or low-energy states of a finite system converge to a single ground state in the infinite volume limit.
In this case, a finite-size scaling analysis of the gap of finite systems does not provide meaningful information about the energy gap in the infinite volume limit.
A typical example is the Affleck-Kennedy-Lieb-Tasaki (AKLT) model on the open chain \cite{AKLT1,AKLT2,TasakiBook}.

Second, when the infinite volume ground states exhibit long-range order in a system where the order operator and the Hamiltonian do not commute, it commonly happens that the finite volume ground state is unique and accompanied by low-lying excited states. See, e.g., Part~I of \cite{TasakiBook}.
In such a situation the spectra of finite systems do not reflect the nature of the infinite volume ground states.
In the antiferromagnetic XXZ chain with strong Ising anisotropy (described by the Hamiltonian \rlb{XXZ} with $H=0$ and $0<J\ll J'$), for example, any finite system has a unique ground state and a low-lying excited state with vanishingly small energy gap while the infinite system has (at least) two symmetry-breaking ground states which are locally-unique and gapped.
}
In an infinite system, on the other hand, one can directly characterize a (locally-)unique gapped ground state as we shall discuss below.\footnote{
It may of course happen that some information of finite systems is lost in the infinite volume limit.
}

It is standard to describe an infinite quantum system not in terms of a state in a Hilbert space, but in terms of the expectation values of local operators.
Such an approach is known as the operator algebraic formulation or the C$^*$-algebraic formulation.
Although a theory based on the operator algebraic formulation may sometimes be mathematically advanced, the basic idea is not difficult and can be understood intuitively.

In section~\ref{s:setting1D}, we discuss in an elementary manner some essential concepts in the operator algebraic treatment of infinite quantum spin chains.
We here introduce the notion of a locally-unique gapped ground state.
In section~\ref{s:finite}, we discuss how a locally-unique gapped ground state is related to unique gapped ground states of finite spin chains.

See Appendix~A.7 of \cite{TasakiBook} for a more detailed introduction to the operator algebraic formulation of quantum spin systems.
The reader interested in full details is invited to study the definitive textbook by Bratteli and Robinson \cite{BR1,BR2}.

\subsection{A locally-unique gapped ground state of a quantum spin system on the infinite chain}
\label{s:setting1D}
Let us formulate a general quantum spin system on the infinite chain $\bbZ$.
We associate each site $j\in\bbZ$ with a quantum spin that has spin quantum number\footnote{
\label{fn:Sj}
We can treat models in which the spin quantum number $S_j$ depends on $j$.
} $S\in\{\frac{1}{2},1,\frac{3}{2},\ldots\}$.
The spin at site $j$ is described by the standard spin operators $\hSx_j$, $\hSy_j$, and $\hSz_j$ on the local Hilbert space $\hilb_j\cong\bbC^{2S+1}$.
As usual, the same symbol $\hSa_j$ (with $\alpha=\mrx$, $\mry$, $\mrz$) denotes the corresponding operator on a larger Hilbert space $\hilb_j\otimes\calH$, which, to be precise, should be written as $\hSa_j\otimes\iop_{\calH}$, where $\calH$ is the space of states external to the site $j$.

We consider a quantum spin system on $\bbZ$ whose Hamiltonian is formally expressed as the infinite sum
\eq
\hH=\sum_{j\in\bbZ}\hh_j,
\lb{H}
\en
where the local Hamiltonian $\hh_j$ is a polynomial of spin operators $\hSa_k$ with $\alpha=\mrx$, $\mry$, $\mrz$ and $k\in\bbZ$ such that $|j-k|\le r_0$.
Here $r_0$ is a constant that determines the range of the interactions.
We also assume that $\norms{\hh_j}\le h_0$ for any $j\in\bbZ$ with a constant $h_0$.\footnote{
Throughout the present article, $\snorm{\hA}$ denotes the operator norm of $\hA\in\Aloc$.
It is defined as $\snorm{\hA}=\sup_{\ket{\Phi}\in\calH\backslash\{0\}}\snorm{\hA\ket{\Phi}}/\snorm{\ket{\Phi}}$, where we regarded $\hA$ as an operator on a suitable finite dimensional Hilbert space $\calH$.
When $\hA$ is self-adjoint, the norm $\snorm{\hA}$ is equal to the maximum of the absolute values of the eigenvalues of $\hA$.
See, e.g., Appendix~A.2 of \cite{TasakiBook} for other properties of the operator norm.
}
The most important example is the antiferromagnetic Heisenberg chain with
\eq
\hH_{\rm HAF}=\sum_{j\in\bbZ}\hat{\boldsymbol{S}}_j\cdot\hat{\boldsymbol{S}}_{j+1},
\lb{HAF}
\en
where $\hat{\boldsymbol{S}}_j=(\hSx_j,\hSy_j,\hSz_j)$.
In this case we may set $\hh_j=\hat{\boldsymbol{S}}_j\cdot\hat{\boldsymbol{S}}_{j+1}$.

We shall now formulate the notion of a locally-unique gapped ground state of an infinite spin chain.
It is standard and convenient to work with the algebra of operators rather than with the Hilbert space.

By a local operator of the spin chain, we mean an arbitrary polynomial of operators $\hSa_j$ with $\alpha=\mrx$, $\mry$, $\mrz$ and $j\in\bbZ$.
The set of sites on which a local operator acts nontrivially is called the support of the operator.
Note that the support of a local operator is always a finite subset of $\bbZ$.
We define the algebra of local operators, which is denoted as $\Aloc$, as the set of all local operators.
It is worth noting that the local Hamiltonian $\hh_j$ belongs to $\Aloc$, but the total Hamiltonian $\hH$ does not.

We can then define the notion of states on the infinite chain as follows.
The idea is that $\rho(\hA)$ is the expectation value of the operator $\hA$ in the state.

\begin{definition}[state]
A state $\rho$ of the spin chain is a liner map from\footnote{%
To be rigorous, a state $\rho$ is defined to be a linear map from the C${}^*$-algebra $\mathfrak{A}$ to $\bbC$, where
$\OA$ is the completion of $\Aloc$ with respect to the operator norm.
} $\Aloc$ to $\bbC$ such that $\rho(\iop)=1$ and $\rho(\hA^\dagger\hA)\ge0$ for any $\hA\in\Aloc$.
\end{definition}
From the definition it follows that $|\rho(\hA)|\le\snorm{\hA}$ and $\rho(\hA^\dagger)=\rho(\hA)^*$ \cite{BR1}.

For an arbitrary local operator $\hV\in\Aloc$, we define its commutator with $\hH$ as
\eq
[\hH,\hV]=\sum_{j\,:\,|j|\le L}[\hh_j,\hV],
\lb{HV}
\en
where $L$ is taken so that the support of $\hV$ is included in $[-L,L]$.
Note that the definition is independent of the choice of $L$.
It is notable that $[\hH,\hV]$ is a well-defined local operator, although $\hH$ is only formally defined by the infinite sum \rlb{H}.
Then a ground state is characterized as follows.

\begin{definition}[ground state]\label{d:GS}
A state $\omega$ is said to be a ground state of $\hH$ if it holds that 
\eq
\omega(\hV^\dagger[\hH,\hV])\ge0,
\lb{GS1}
\en
for any $\hV\in\Aloc$.
\end{definition}
In short, the definition says that one can never lower the energy expectation value of the state $\omega$ by perturbing it with a local operator $\hV$.
To see this more clearly, consider a finite system and let $\omega$ be a state written as $\omega(\cdot)=\bra{\Phi}\cdot\ket{\Phi}$ with a pure normalized state $\ket{\Phi}$.
Then the condition \rlb{GS1} reads 
\eq
\bra{\Phi}\hV^\dagger\hH\hV\ket{\Phi}\ge\bra{\Phi}\hV^\dagger\hV\hH\ket{\Phi}.
\lb{GS1B}
\en
If we take $\ket{\Phi}$ as a ground state $\GS$ with energy $E_{\rm GS}$, this is rewritten as $\bra{\Psi}\hH\ket{\Psi}\ge E_{\rm GS}$, 
where 
\eq
\ket{\Psi}=\frac{\hV\GS}{\snorm{\hV\GS}}
\lb{Psi}
\en
is a normalized variational state.
We thus get the standard variational characterization of a ground state.
When $\ket{\Phi}$ is not a ground state one immediately sees that \rlb{GS1B} is violated by taking $\hV=\GS\bra{\Phi}$.

We now state the notion of a locally-unique gapped ground state, which is central to the present article.

\begin{definition}[locally-unique gapped ground state]\label{d:LUGS}
A ground state $\omega$ of $\hH$ is said to be a locally-unique gapped ground state if there is a constant $\gamma>0$ such that
\eq
\omega(\hV^\dagger[\hH,\hV])\ge\gamma\,\omega(\hV^\dagger\hV),
\lb{GS2}
\en
holds for any $\hV\in\Aloc$ with $\omega(\hV)=0$.
The energy gap $\DE$ of the ground state $\omega$ is the largest $\gamma$ with the above property.
\end{definition}
To see the relation with the standard definition for a finite system, observe that the conditions read $\bra{\Psi}\hH\ket{\Psi}\ge E_{\rm GS}+\gamma$ and $\GSb\Psi\rangle=0$ for the same $\GS$ and $\ket{\Psi}$ as above.
These are precisely the variational characterization of a unique gapped ground state.

The main part of the following important theorem that characterizes a locally unique gapped ground state was proved by Matsui \cite{Matsui2010,Matsui2013}, who made use of the result of Hastings \cite{Hastings2007}.
The present version that only assumes local-uniqueness is due to Ogata.
See Appendix~\ref{s:Yoshiko}.
We note that the theorem is valid only in one dimension (while all the definitions in the present section and Theorem~\ref{t:finiteGS} in the next section readily extend to higher dimensions).
\begin{theorem}\label{t:puresplit}
A locally-unique gapped ground state is a pure split state.
\end{theorem}
The reader does not have to understand the precise definition of a pure split state in order to go through the present article.
Roughly speaking, a state is said to be pure if it cannot be written as a mixed state, i.e., a convex sum of more than one state.
A state is said to satisfy the split property if it has small entanglement, or, more precisely, its entanglement entropy obeys the area law.

As the terminology ``locally-unique gapped ground state" suggests, the above definition is different from the following standard definition of a unique gapped ground state.
\begin{definition}[unique gapped ground state]\label{d:UGS}
A ground state $\omega$ of $\hH$ is said to be a unique gapped ground state if it is the only state that satisfies the conditions of Definition~\ref{d:GS}, and there is a constant $\gamma>0$ such that $\omega(\hV^\dagger[\hH,\hV])\ge\gamma\,\omega(\hV^\dagger\hV)$ holds for any $\hV\in\Aloc$ with $\omega(\hV)=0$.
\end{definition}
The difference is that Definition~\ref{d:LUGS} only requires the uniqueness within the space of states that can be reached from $\omega$ by local perturbations, while Definition~\ref{d:UGS} requires much stronger global uniqueness.
The two definitions coincide in a finite system, but they are different in an infinite system.

A unique gapped ground state is clearly a locally-unique gapped ground state, but the converse may not be true.
When there is a spontaneous symmetry breaking, one may have a locally-unique gapped ground state that is not a unique gapped ground state.
As a simple example, consider the $S=1/2$ ferromagnetic Ising model with the Hamiltonian
\eq
\hH_{\rm Ising}=-\sum_{j\in\bbZ}\hSz_j\hSz_{j+1}.
\lb{ferroIsing}
\en
The all-up state, formally written as $\bigotimes_{j\in\bbZ}\ket{\!\!\up}_j$, is a locally-unique gapped ground state since any local modification of the configuration leads to an increase in the energy.
But it is not a unique ground state since the all-down state is also a locally-unique gapped ground state\footnote{%
The domain wall states $(\bigotimes_{j:j\le k}\ket{\!\!\up}_j)\otimes(\bigotimes_{j:j> k}\ket{\!\!\downarrow}_j)$ and $(\bigotimes_{j:j\le k}\ket{\!\!\downarrow}_j)\otimes(\bigotimes_{j:j> k}\ket{\!\!\up}_j)$ with any $k$ are also ground states of the model.
These ground states are of course not locally-unique.
}.
See foonote~\ref{fn:toric} in Apendix~\ref{s:Yoshiko} for an interesting example in two dimensions related to the topological order of a locally-unique gapped ground state that is not a unique ground state.

%%%%%%%%%
\subsection{Relation to unique gapped ground states on finite chains}
\label{s:finite}
Let us see how the above rather abstract definition of a locally-unique gapped ground state is related to the definition that is standard in physics.
Although the result is valid in any dimension, we here restrict ourselves to one-dimensional systems for simplicity.
In what follows, $\Aloc$ and $\hH$ denote exactly the same objects as in section~\ref{s:setting1D}.

Let $L$ be an even integer, and consider a quantum spin system on the finite chain $\CL=\{-L/2,\ldots,L/2\}\subset\bbZ$ with the Hamiltonian $\hH_L$ written as
\eq
\hH_L=\sum_{j\in{\mathcal{I}}_L}\hh_j+\Di\hH_L,
\en
where ${\mathcal{I}}_L\subset\CL$ is the set of $j$ such that the support of $\hh_j$ is contained in $\CL$.
The local Hamiltonian $\hh_j$ is the same as in \rlb{H}.
Here $\Di\hH_L$ is a suitably chosen boundary Hamiltonian that acts on sites within a fixed distance (independent of $L$) from the two boundaries.
One can consider a Hamiltonian with the periodic boundary condition as a special case.

We assume that, for each $L$, the finite volume Hamiltonian $\hH_L$ has a unique normalized ground state $\ket{\Phi_{\rm GS}^{(L)}}$ accompanied by a nonzero energy gap that is not less than a constant $\DE>0$.
Note that we are using the standard physicists' notation since this is only quantum mechanics with a finite-dimensional Hilbert space.

We then define a state $\omega$ on the infinite chain by
\eq
\omega(\hA)=\lim_{L\up\infty}\bra{\Phi_{\rm GS}^{(L)}}\hA\ket{\Phi_{\rm GS}^{(L)}},
\lb{lim1}
\en
for any $\hA\in\Aloc$.  
Note that the expectation value $\bra{\Phi_{\rm GS}^{(L)}}\hA\ket{\Phi_{\rm GS}^{(L)}}$ is well-defined for sufficiently large $L$ since $\hA$ is local.
Of course the limit \rlb{lim1} may not exist.
It is known however that one can always take a subsequence, i.e., a strictly increasing function $L(n)\in\bbN$ of $n\in\bbN$, such that the infinite volume limit
\eq
\omega(\hA)=\lim_{n\up\infty}\bra{\Phi_{\rm GS}^{(L(n))}}\hA\ket{\Phi_{\rm GS}^{(L(n))}},
\lb{lim2}
\en
exists for any $\hA\in\Aloc$.\footnote{
This is an abstract statement and applies to any infinite sequence of states.
Mathematically, this is the Banach-Alaoglu theorem, and the limit is known as the weak-$*$ limit.}
This defines a state of the infinite chain, although the limit may not be unique in general.

The infinite volume state $\omega$ is a ground state but not necessarily a unique gapped ground state.\footnote{
A simple example is the all-up state, which is obtained as the limit of the ground states of the Ising model \rlb{ferroIsing} with the $+$ boundary conditions.
}
But we see that it is always a locally-unique gapped ground state.

\begin{theorem}
\label{t:finiteGS}
The limiting infinite volume state $\omega$ is a locally-unique gapped ground state of $\hH$.
\end{theorem}

\noindent
{\em Proof:}
We write $\omega_L(\cdot)=\bra{\Phi_{\rm GS}^{(L)}}\cdot\ket{\Phi_{\rm GS}^{(L)}}$.
Let $\hV\in\Aloc$ be an arbitrary local operator.
Take sufficiently large $L$ such that the support of $\hV$ is contained in $\CL$ and does not overlap with the support of $\Di\hH_L$.
We then have $[\hH_L,\hV]=[\hH,\hV]$.

On the other hand the standard variational principle implies $\omega_L(\hV^\dagger\,[\hH_L,\hV])\ge0$, which means $\omega_L(\hV^\dagger\,[\hH,\hV])\ge0$.
By letting $L\up\infty$ (or, more precisely, $n\up\infty$ in $L(n)$) we see that $\omega$ is a ground state in the sense of Definition~\ref{d:GS}.

To see that $\omega$ is locally-unique and gapped, we further assume that $\omega(\hV)=0$.
Let $\hV_L=\hV-\omega_L(\hV)$.
Note that this means $\hV_L\ket{\Phi_{\rm GS}^{(L)}}$ is orthogonal to $\ket{\Phi_{\rm GS}^{(L)}}$.
Since $\ket{\Phi_{\rm GS}^{(L)}}$ is unique and gapped, we find from the variational principle that
\eq
\bra{\Phi_{\rm GS}^{(L)}}\hV_L^\dagger\hH_L\hV_L\ket{\Phi_{\rm GS}^{(L)}}
\ge
(E_{\rm GS}^{(L)}+\DE)\,\bra{\Phi_{\rm GS}^{(L)}}\hV_L^\dagger\hV_L\ket{\Phi_{\rm GS}^{(L)}},
\en
which is rewritten as
\eq
\omega_L(\hV_L^\dagger\,[\hH_L,\hV_L])\ge\DE\,\omega_L(\hV_L^\dagger\,\hV_L).
\en
This means, for sufficiently large $L$, that
\eq
\omega_L(\hV_L^\dagger\,[\hH,\hV])\ge\DE\,\omega_L(\hV_L^\dagger\,\hV_L).
\en
Since $\hV_L\to\hV$ as $L\up\infty$, we get the desired condition
\eq
\omega(\hV^\dagger\,[\hH,\hV])\ge\DE\,\omega(\hV^\dagger\,\hV),
\en
for the infinite volume ground state.~\qedm

\medskip

This important theorem states that if one has a sequence of finite chains with a unique gapped ground state then the corresponding state in the infinite volume limit is necessarily a locally-unique gapped ground state.
We see that the notion of a locally-unique gapped ground state is natural from the physical point of view.

In the following sections, we shall discuss LSM-type theorems that rule out the possibility of a locally-unique gapped ground state in the infinite chain.
Thanks to Theorem~\ref{t:finiteGS}, such an abstract no-go theorem implies that one is never able to choose a sequence of finite volume Hamiltonians $\hH_L$ with a unique gapped ground state.

%%%%%%%%%%%%%%%%%%%%%%%%
\section{Generalized LSM theorem for translation-invariant spin chains with U(1) symmetry}
\label{s:U1}
In the present section, we focus on translation-invariant quantum spin chains with U(1) symmetry and discuss a generalized LSM theorem that can be proved by the philosophy in the original paper of Lieb, Schultz, and Mattis \cite{LSM}.

\medskip

In Appendix~B of their paper in 1961, Lieb, Schultz, and Mattis stated their no-go theorem for the  $S=1/2$ antiferromagnetic model with Hamiltonian \rlb{HAF} on the finite periodic chain with an even number of sites \cite{LSM}.
In this case one knows from the Marshall-Lieb-Mattis theorem \cite{Marshall,LiebMattis} (which is indeed Lemma~2 in Appendix~B of \cite{LSM}) that the ground state is unique and has vanishing total spin.
By using the (global)  twist unitary operator \rlb{Utwist} they constructed a trial state with low excitation energy, and proved that the energy gap immediately above the ground state in this model is bounded by a constant times $L^{-1}$, where $L$ denotes the length of the chain.

In 1986, Affleck and Lieb made an essential generalization of the theorem \cite{AL}.
They studied general translation-invariant quantum spin chains with $\Uo\rtimes\bbZ_2$ symmetry, where U(1) corresponds to the rotation about the z-axis by an arbitrary amount and $\bbZ_2$ the $\pi$-rotation about the x-axis.\footnote{
Let $g_\theta$ denote the $\theta$-rotation about the z-axis and $r$ the $\pi$-rotation about the x-axis.
(One may think about spatial rotations.)
Noting that $r\circ g_\theta\circ r=g_{-\theta}$, we see that the whole group is not the direct procut $\Uo\times\bbZ_2$ but the semidirect product $\Uo\rtimes\bbZ_2$.
In general, $G\rtimes H$ or $H\ltimes G$ denotes the semidirect product of $G$ and $H$ where $H$ acts on $G$.
}
A typical examples is the XXZ model \rlb{XXZ} with $H=0$.
They found that the same conclusion as the original LSM theorem can be derived when and only when the spin quantum number $S$ is a half-odd-integer.
This finding is consistent with Haldane's conclusion that the spin $S$ Heisenberg antiferromagnetic chain \rlb{HAF} has a unique gapped ground state if $S$ is an integer and a unique gapless ground state if $S$ is a half-odd-integer \cite{Haldane1981,Haldane1983a,Haldane1983b}.
Note that the Affleck-Lieb theorem rigorously justifies the latter half of Haldane's conclusion.
The first half of the conclusion is still unproven, but there are strong indications, including a rigorous example by Affleck, Kennedy, Lieb, and Tasaki \cite{AKLT1,AKLT2}, that it is correct.
See, e.g., part~II of \cite{TasakiBook} for more about this fascinating topic.

In the same paper \cite{AL}, Affleck and Lieb also formulated and proved their generalized LSM theorem for infinite quantum spin chains.
As we have discussed at the beginning of section~\ref{s:Gen}, a no-go theorem that rules out a unique gapped ground state should ultimately be stated for the infinite chain.
See also footnote~\ref{fn:finite}.
In this sense the extension to infinite systems in this work is essential.
In the present article, we follow the formulation of Affleck and Lieb and discuss LSM-type theorems for infinite systems.

In 1997, Oshikawa, Yamanaka, and Affleck \cite{OYA} made a further generalization, which revealed the meaning of the restriction of the Affleck-Lieb theorem to spin chains with half-odd-integral spins.
They studied a general translation-invariant quantum spin chain with only $\Uo$ symmetry, such as the XXZ model \rlb{XXZ} under nonzero magnetic field $H$.
By extending the method of the original work by Lieb, Schultz, and Mattis \cite{LSM}, they found that the spin chain cannot have a unique gapped ground state if $\nu=\bra{\Phi_{\rm GS}}\hSz_j\GS+S$ is not an integer.
Note that one has $\nu=S$ in a $\Uo\rtimes\bbZ_2$ symmetric ground state, and hence the condition reduces to the requirement that $S$ is a half-odd-integer.
The quantity $\nu$ is called the filling factor and is now regarded as central in the discussion of generalized LSM theorems in U(1) invariant quantum systems.
In \cite{OYA}, quantum spin systems on finite periodic chains were treated.
The corresponding generalized LSM theorem for infinite spin chains was proved later in \cite{TasakiLSM}.

\medskip

In the present section, we give a full proof of a generalized LSM theorem for infinite spin chains that unifies the results of Lieb, Schultz, and Mattis \cite{LSM}, Affleck and Lieb \cite{AL}, and Oshikawa, Yamanaka, and Affleck \cite{OYA}.
The proof is essentially a rearrangement of that in \cite{TasakiLSM}.
We again make essential use of the local twist operator introduced by Affleck and Lieb \cite{AL}, but in a slightly different way, namely, to define a $\bbZ$-valued topological index for a locally-unique gapped ground state.
By identifying the index with the filling factor, we arrive at a generalized LSM theorem.
We believe that the new proof is not only simpler than that in \cite{TasakiLSM} but is also interesting in its own light.\footnote{
In \cite{TasakiLSM}, a statement about the number of low-energy excited states in a finite chain is also proved.
See Corollary 2.
We are unable to reproduce such detailed information in the present simpler proof that directly works for the infinite chain.
}

As we noted in section~\ref{s:intro}, a generalized LSM theorem for quantum spin systems in any dimension was proved by Bachmann, Bols, De Roeck, and Fraas \cite{BachmannBolsDeRoecFraas2019} by using a similar strategy.
Our proof, which is restricted to one dimension, is more elementary.

\subsection{The local twist operator and its expectation value}
We here follow the formulation in section~\ref{s:Gen} and study a quantum spin system on the infinite chain $\bbZ$.

Throughout the present section we only treat Hamiltonians that have U(1) invariance, or, more precisely, is invariant under any uniform spin-rotation about the z-axis.
To make the assumption precise, let us denote by
\eq
\hU^I_\theta=\exp[-i\sum_{j\in I}\theta\hSz_j],
\lb{UI}
\en
the unitary operator for the $\theta$-rotation about the z-axis of spins in a finite interval $I\subset \bbZ$.
We then assume for any $j\in\bbZ$ that
\eq
(\hU^I_\theta)^\dagger\hh_j\hU^I_\theta=\hh_j,
\lb{U1}
\en
for any $\theta\in[0,2\pi]$, where $I$ is an arbitrary interval that contains the support of $\hh_j$.

A typical example is the XXZ model under uniform magnetic field with the Hamiltonian
\eq
\hH_{\rm XXZ}=\sum_{j\in\bbZ}\bigl\{J(\hSx_j\hSx_{j+1}+\hSy_j\hSy_{j+1})+J'\hSz_j\hSz_{j+1}-H\hSz_j\bigr\},
\lb{XXZ}
\en
where $J,J'\in\bbR$ are anisotropic exchange interaction constants and $H\in\bbR$ is the magnetic field.

\begin{figure}
\centerline{\epsfig{file=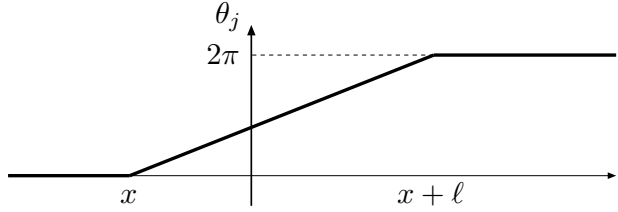,width=8truecm}}
\caption[dummy]{
The spin-rotation angle $\theta_j$ increases gradually from 0 to $2\pi$ as the coordinate varies from $x$ to $x+\ell$.  Otherwise it is equal to 0 or $2\pi$.
}
\label{f:theta}
\end{figure}

Let us note that the (global) twist operator $\hU_{\rm twist}$ of Bloch \cite{Bohm,TadaKoma,Watanabe2019} and Lieb, Schultz, and Mattis \cite{LSM} defined in \rlb{Utwist} does not make sense in the infinite volume limit.
We follow Affleck and Lieb \cite{AL} and introduce the local twist operator, which plays a central role in the present section and section~\ref{s:2D}.
For any $x\in\bbR$ and $\ell>0$, we introuduce the corresponding angle $\theta_j$ for $j\in\bbZ$ by
\eq
\theta_j=\begin{cases}
0,&j< x;\\
2\pi(j-x)/\ell,&x\le j\le x+\ell;\\
2\pi,&j>x+\ell,
\end{cases}
\lb{theta}
\en
and define the local twist operator by
\eq
\hUx=\exp\Bigl[-i\sum_{j\in\bbZ}\theta_j(\hSz_j+S)\Bigr].
\lb{Uxl}
\en
Although the definition has an infinite sum inside the exponential function, it is indeed a local operator because one has $\exp[-i2\pi(\hSz_j+S)]=1$.
In fact we can write the same operator as
\eq
\hUx=\exp\Bigl[-i\sum_{j\in\bbZ\cap[x,x+\ell]}\frac{2\pi(j-x)}{\ell}(\hSz_j+S)\Bigr],
\lb{Uxl2}
\en
where the locality is manifest.
Like the operator \rlb{UI}, the twist operator $\hUx$ describes a spin-rotation in the interval $\bbZ\cap[x,x+\ell]$, but the rotation angle $2\pi(j-x)/\ell$ varies gradually from 0 to $2\pi$ as $j$ moves from $x$ to $x+\ell$.
When $\ell$ is large, the operator gives a gentle twist to the state on which it acts.
See Figure~\ref{f:theta}.

Note that the twist operator $\hUx$ is continuous both in $x$ and $\ell$ because  of the identity $\exp[-i2\pi(\hSz_j+S)]=1$.
The continuity will play an essential role in the present analysis.
We used $\hSz_j+S$, instead of $\hSz_j$, in the definition \rlb{Uxl} to guarantee the continuity\footnote{This is only necessary when $S$ is a half-odd-integer.}.

We shall start by stating an elementary but essential variational estimate that goes back to Lieb, Schultz, and Mattis \cite{LSM}.

\begin{lemma}\label{l:var}
Let $\omega$ be a ground state of an $\rm U(1)$ invariant Hamiltonian $\hH$.
Then it holds for any $x\in\bbR$ and $\ell\ge\ell_0$ that
\eq
0\le\omega(\hUx^\dagger\,[\hH,\hUx])\le\frac{C}{\ell},
\lb{var}
\en
where $C$ and $\ell_0$ are constants that depend on the Hamiltonian.\footnote{%
One may express $C$ in terms of $S$, $r$, and $h_0$.
See \rlb{varex} below for an example.
}
\end{lemma}

The first inequality in \rlb{var} is nothing but \rlb{GS1} in the definition of ground states.
Consider, as we did in the interpretation of the condition \rlb{GS1}, a finite system and let $\omega(\hA)=\bra{\Phi_{\rm GS}}\hA\GS$, where $\GS$ is a normalized ground state with energy $E_{\rm GS}$.
Then we see that
\eq
\omega(\hUx^\dagger\,[\hH,\hUx])=\bra{\Phi_{\rm GS}}\hUx^\dagger\hH\hUx\GS-\bra{\Phi_{\rm GS}}\hH\GS
=\bra{\Psi}\hH\ket{\Psi}-E_{\rm GS},
\en
where $\ket{\Psi}=\hUx\GS$ is a noramalized trial state.
Thus we see that the expectation value $\omega(\hUx^\dagger\,[\hH,\hUx])$ is the increase in the energy expectation value when the ground state $\omega$ is modified by a local unitary operator $\hUx$.
The lemma states that the increase can be made as small as one wishes by letting $\ell$ large.

\medskip
\noindent
{\em Proof:}
Since \rlb{GS1} also implies $\omega(\hUx\,[\hH,\hUx^\dagger])\ge0$ for a ground state $\omega$, we have
\eq
\omega(\hUx^\dagger\,[\hH,\hUx])\le
\omega(\hUx^\dagger\,[\hH,\hUx])+\omega(\hUx\,[\hH,\hUx^\dagger])
=\omega([\hUx^\dagger\,,[\hH,\hUx]]),
\lb{omega1}
\en
where the final expression in terms of a double commutator can be easily confirmed by an explicit computation.

Let $I$ be the set of $j$ such that the support of $\hh_j$ overlaps with the interval $[x,x+\ell]$.
Noting that $[\hh_j,\hUx]=0$ for $j\not\in I$, we see that
\eq
[\hUx^\dagger\,,[\hH,\hUx]]=\sum_{j\in I}[\hUx^\dagger\,,[\hh_j,\hUx]]
=\sum_{j\in I}\{\hUx^\dagger\,\hh_j\hUx+\hUx\,\hh_j\hUx^\dagger-2\hh_j\}.
\lb{UHU}
\en
Recalling the assumed U(1) invariance \rlb{U1}, we expect that the summand in the final expression should be small since $\hUx$ is locally close to the uniform rotation $\hU_\theta^I$.

For simplicity let us focus on the XXZ model with \rlb{XXZ}, and write the Hamiltonian as $\hH=\sum_j\hh_j$ with $\hh_j=\hh^{\rm XY}_j+\hh^{\rm Z}_j$, where 
\eq
\hh^{\rm XY}_j=J(\hSx_j\hSx_{j+1}+\hSy_j\hSy_{j+1})=\frac{J}{2}(\hS^+_j\hS^{-}_{j+1}+\hS^-_j\hS^{+}_{j+1}),
\en
and $\hh^{\rm Z}_j=J'\hSz_j\hSz_{j+1}-H\hSz_j$, where we defined $\hS_j^\pm=\hSx_j\pm i\hSy_j$.
We then find from an explicit computation that
\eqa
\hUx^\dagger\,\hh_j\hUx+\hUx\,\hh_j\hUx^\dagger-2\hh_j
&=\hUx^\dagger\,\hh^{\rm XY}_j\hUx+\hUx\,\hh^{\rm XY}_j\hUx^\dagger-2\hh^{\rm XY}_j
\nl
&=\frac{J}{2}\bigl\{e^{i(\theta_j-\theta_{j+1})}\hS^+_j\hS^{-}_{j+1}+e^{-i(\theta_j-\theta_{j+1})}\hS^+_j\hS^{-}_{j+1}-2\hS^+_j\hS^{-}_{j+1}
\nl&\hspace{19pt}+e^{-i(\theta_j-\theta_{j+1})}\hS^-_j\hS^{+}_{j+1}+e^{i(\theta_j-\theta_{j+1})}\hS^-_j\hS^{+}_{j+1}-2\hS^-_j\hS^{+}_{j+1}\bigr\}
\nl&=2J\bigl\{\cos(\theta_j-\theta_{j+1})-1\bigr\}(\hSx_j\hSx_{j+1}+\hSy_j\hSy_{j+1}),
\lb{UhU}
\ena
where $\theta_j$ is given in \rlb{theta}.
Noting that $0\le1-\cos\theta\le\theta^2/2$, the norm of \rlb{UhU} is bounded as
\eq
\snorm{\hUx^\dagger\,\hh_j\hUx+\hUx\,\hh_j\hUx^\dagger-2\hh_j}
\le2J(\theta_j-\theta_{j+1})^2S^2\le\frac{8\pi^2S^2}{\ell^2}.
\lb{hUh2}
\en
Since the interval $I$ may contain at most $\ell+1$ sites, we get from \rlb{omega1}, \rlb{UHU}, and \rlb{hUh2} that
\eq
\omega(\hUx^\dagger\,[\hH,\hUx])\le\Bigl\Vert[\hUx^\dagger\,,[\hH,\hUx]]\Bigr\Vert\le8\pi^2S^2\frac{\ell+1}{\ell^2}\le\frac{16\pi^2S^2}{\ell},
\lb{varex}
\en
where the final inequality is valid for $\ell\ge\ell_0=1$.

A general U(1) invariant Hamiltonian can be treated similarly.
See, e.g., section~6.2 of \cite{TasakiBook}.~\qedm

\medskip
The next lemma about the expectation value of the twist operator is the main result of the present subsection.

\begin{lemma}
Let $\omega$ be a locally-unique gapped ground state with gap $\DE>0$ of a $\rm U(1)$ invariant Hamiltonian $\hH$.
We then have
\eq
|\omega(\hUx)|^2\ge1-\frac{C}{\ell}\frac{1}{\DE},
\lb{wU>}
\en
for any $x\in\bbR$ and $\ell\ge\ell_0$, where the constants $C$ and $\ell_0$ are the same as in Lemma~\ref{l:var}.
\end{lemma}

Since $|\omega(\hUx)|\le1$, the inequality implies that $|\omega(\hUx)|$ tends to unity as $\ell\up\infty$.
This result is intuitively clear if one notes that the corresponding conclusion in a finite system is $\bigl|\GSb\Psi\rangle\bigr|\simeq1$ for $\ket{\Psi}=\hUx\GS$.
Recall that the unitary operator $\hUx$, when acting on a ground state, increases the energy expectation value by only a small amount.
Since the ground state is locally-unique and gapped, the only state with such low energy is the ground state itself.

\medskip
\noindent
{\em Proof:}
Let $\hW=\hUx-\omega(\hU_x)$.
Since $\omega$ is a locally-unique gapped ground sate and $\omega(\hW)=0$, we have
\eq
\omega(\hW^\dagger[\hH,\hW])\ge\DE\,\omega(\hW^\dagger\hW),
\en 
which is nothing but \rlb{GS2}.
Substituting the definition of $\hW$ and noting that\footnote{
If $\omega$ is a ground state of $\hH$, we have $\omega([\hH,\hA])=0$ for any $\hA\in\Aloc$.
To see this, note that \rlb{GS1} implies that $\omega(\hV^\dagger\hV\hH)\in\bbR$.
We then have $\omega(\hV^\dagger\hV\hH)=\omega((\hV^\dagger\hV\hH)^\dagger)=\omega(\hH\hV^\dagger\hV)$, and hence $\omega([\hH,\hV^\dagger\hV])=0$.
Since any local operator is written as $\hA=(\hV_1^\dagger\hV_1-a)+i(\hV_2^\dagger\hV_2-b)$ with suitable $\hV_1,\hV_2\in\Aloc$ and $a,b\in\bbR$, the claim has been proved.
} $\omega([\hH,\hW])=0$, we see that the above inequality reduces to
\eq
\omega(\hUx^\dagger[\hH,\hUx])\ge\DE\bigl(1-|\omega(\hUx)|^2\bigr).
\en
Then the variational estimate \rlb{var} implies the desired \rlb{wU>}.~\qedm

\subsection{Generalized LSM theorem}
\label{s:U1LSM}
Let $\omega$ be a locally-unique gapped ground state with energy gap $\DE>0$ of a U(1) invariant Hamiltonian $\hH$ on $\bbZ$.
We further assume that $\omega$ is invariant under translation by $p$, where $p$ is a positive integer, in the sense that 
\eq
\omega(\calT_p(\hA))=\omega(\hA),
\lb{TI}
\en
for any local operator $\hA$.
Here $\calT_p$ denotes the transformation (the linear $*$-automorphism\footnote{
\label{fn:*auto}
In general a linear  $*$-automorphism is a one-to-one linear map $\Gamma:\Aloc\to\Aloc$ that satisfies 
$\Gamma(\hA\hB)=\Gamma(\hA)\Gamma(\hB)$ and 
$\Gamma(\hA^\dagger)=\Gamma(\hA)^\dagger$
for any $\hA,\hB\in\Aloc$.
In a finite quantum system, a linear $*$-automorphism $\Gamma$ is always written as
$\Gamma(\hA)=\hU^\dagger\hA\hU$ for any $\hA\in\Aloc$ with a suitable unitary operator $\hU$.  See, e.g., Appendix~A.6 of \cite{TasakiBook}.
}) such that $\calT_p(\hSa_j)=\hSa_{j+p}$ for any $\alpha=\rm x,y,z$ and $j\in\bbZ$.
It also satisfies $\calT_p(\hA\hB)=\calT_p(\hA)\calT_p(\hB)$ and $\calT_p(\hA^\dagger)=\{\calT_p(\hA)\}^\dagger$ for any  $\hA,\hB\in\Aloc$.

We here do not assume the translation invariance of the Hamiltonian $\hH$ since we do not make use of that assumption.
We note however that one usually gets a translation-invariant ground state as a (locally-)unique ground state of a translation-invariant Hamiltonian.

\begin{figure}
\centerline{\epsfig{file=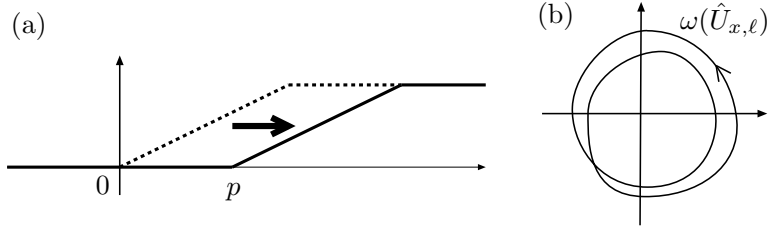,width=10truecm}}
\caption[dummy]{
(a) When $x$ is varied from $0$ to $p$, the profile (see Figure~\ref{f:theta}) of the rotation angle is modified continuously.
(b) Since $\omega(\hU_{0,\ell})=\omega(\hU_{p,\ell})$, the expectation value $\omega(\hUx)\ne0$ for $x\in[0,p]$ defines a closed oriented path in the complex plane that does not touch the origin.
One can then determine the winding number that characterizes the path or the ground state $\omega$.
The winding number is two in this figure.
}
\label{f:winding}
\end{figure}

Take any $\ell$ larger than $\ell_1=\max\{C/\DE,\ell_0\}$.
Then \rlb{wU>} guarantees that $\omega(\hUx)\ne0$ for any $x$.
Since the translation invariance \rlb{TI} implies 
\eq
\omega(\hU_{0,\ell})=\omega(\hU_{p,\ell}), 
\lb{oUoU}
\en
and $\omega(\hUx)$ is continuous in $x$, we see that $\omega(\hUx)$ defines a closed oriented path in $\bbC\backslash\{0\}$ when $x$ is varied from $0$ to $p$ with $\ell$ fixed.
We can then define the winding number of the path around the origin, which we denote as $n_\omega\in\bbZ$.
See Figure~\ref{f:winding}.
Since $\omega(\hUx)$ is continuous also in $\ell$, we see that the winding number $n_\omega$ is independent of the choice of $\ell$, as long as it is larger than $\ell_1$.
Although we do not use this property in the present article, let us also note that the winding number $n_\omega$ is a topological index of the ground state, i.e., it is invariant under continuous modification of locally-unique gapped ground states.\footnote{
Consider a family of U(1) invariant Hamiltonians $\hH_s$ parameterized by $s\in[0,1]$, and assume that $\hH_s$ has a locally-unique gapped ground state $\omega_s$ with gap not smaller than a constant $\DE>0$ independent of $s$.
We further assume that $\omega_s(\hA)$ is continuous in $s$ for any $\hA\in\Aloc$.
One then readily finds that $n_{\omega_s}$ is independent of $s$.
}

For a ground state invariant under translation by $p$, we follow Oshikawa, Yamanaka, and Affleck \cite{OYA} and define its filling factor as
\eq
\nu_\omega=\omega\Bigl(\sum_{j=1}^p(\hSz_j+S)\Bigr).
\lb{nu}
\en
Of course $\omega(\sum_{j=1}^p\hSz_j)$ is the total magnetization in the unit cell $\{1,\ldots,\ell\}$.
We call $\nu_\omega$ the filling factor with the standard mapping of a spin system to a boson system in mind.
More precisely we interpret the state $\ket{\sigma}_j$ such that $\hSz_j\ket{\sigma}_j=\sigma\ket{\sigma}_j$ with $\sigma\in\{-S,-S+1,\ldots,S\}$ as a state in which there are $\sigma+S\in\{0,1,\ldots,2S\}$ particles on site $j$.
Then $\nu_\omega$ is the average of the particle number in the unit cell.

An essential observation is the following index theorem that asserts that the winding number is identical to the filling factor.
We note that the theorem is implicit in the proofs in \cite{LSM,AL,OYA,TasakiLSM} of the orthogonality of the trial states (such as $\ket{\Psi}$ in \rlb{PsiLSM}) to the ground state.

\begin{lemma}
Let $\omega$ be a locally-unique gapped ground state of a $\rm U(1)$ invariant Hamiltonian $\hH$.
We further assume that $\omega$ is invariant under translation by $p$.
Then one has $n_\omega=\nu_\omega$.
\end{lemma}

\noindent
{\em Proof:}
For notational simplicity we only treat the case with $p=1$.
Assuming that $\ell$ is an integer, we have for $x\in[0,1]$ that
\eqa
\hUx&=\exp\Bigl[-i\sum_{j=1}^\ell\frac{2\pi(j-x)}{\ell}(\hSz_j+S)\Bigr]
=\exp\Bigl[i2\pi x\frac{1}{\ell}\sum_{j=1}^\ell(\hSz_j+S)\Bigr]\,\exp\Bigl[-i\sum_{j=1}^\ell\frac{2\pi j}{\ell}(\hSz_j+S)\Bigr]\nl&=e^{i2\pi x\,\hat{\nu}_\ell}\,\hU_{0,\ell},
\lb{UeU}
\ena
with
\eq
\hat{\nu}_\ell=\frac{1}{\ell}\sum_{j=1}^\ell(\hSz_j+S).
\en
From the translation invariance we have $\nu_\omega=\omega(\hat{\nu}_\ell)$ for any positive integer $\ell$. 
Taking the expectation value of \rlb{UeU}, we see $\omega(\hUx)=\omega(e^{i2\pi x\,\hat{\nu}_\ell}\,\hU_{0,\ell})$.
Observe that, if one can replace $\hat{\nu}_\ell$ in the right-hand side with its expectation value $\nu_\omega$, then we get
\eq
\omega(\hUx)\simeq e^{i2\pi x\,\nu_\omega}\,\omega(\hU_{0,\ell}),
\lb{oo2}
\en
which shows that the winding number is $\nu_\omega$.
We shall make this argument rigorous below.

To this end we note that \rlb{UeU} implies
\eqa
\bigl|\omega(\hUx)-e^{i2\pi x\,\nu_\omega}\,\omega(\hU_{0,\ell})\bigr|
&=\Bigl|\omega\bigl((e^{i2\pi x\,\hat{\nu}_\ell}-e^{i2\pi x\,\nu_\omega})\,\hU_{0,\ell}\bigr)\Bigr|
\nl&\le\omega\bigl((e^{-i2\pi x\,\hat{\nu}_\ell}-e^{-i2\pi x\,\nu_\omega})(e^{i2\pi x\,\hat{\nu}_\ell}-e^{i2\pi x\,\nu_\omega})\bigr)\,\omega(\hU^\dagger_{0,\ell}\,\hU_{0,\ell})
\nl&=4\,\omega\bigl(\bigl\{\sin[\pi x (\hat{\nu}_\ell-\nu_\omega)]\bigr\}^2\bigr)
\nl&\le 4\pi^2x^2\,\omega((\hat{\nu}_\ell-\nu_\omega)^2),
\lb{oo}
\ena
where we used the Schwarz inequality to get the upper bound in the second line.
Note that $\omega((\hat{\nu}_\ell-\nu_\omega)^2)$ is the variance of the magnetization (or particle) density $\hat{\nu}_\ell$, and should vanish as $\ell$ grows unless the state is pathological.
To be more precise, noting that
\eq
\omega((\hat{\nu}_\ell-\nu_\omega)^2)=\frac{1}{\ell^2}\sum_{j,k=1}^\ell\omega\Bigl(\bigl(\hSz_j-\omega(\hSz_j)\bigr)\bigl(\hSz_k-\omega(\hSz_k)\bigr)\Bigr),
\en
we see that the quantity in the left-hand side vanishes as $\ell\up\infty$ if the truncated two-point correlation function in the summand has the clustering property, i.e., it vanishes as $|j-k|\up\infty$.
Since we know from Theorem~\ref{t:puresplit} that $\omega$ is a pure state, the general theory \cite{BR1} guarantees the clustering.
Alternatively one can invoke the exponential clustering theorems for gapped ground states proved in \cite{HastingsKoma,NS1} when suitable conditions are met.

We thus conclude that the right-hand side of \rlb{oo} can be made as small as one wishes by letting $\ell$ large.
Recalling also that $|\omega(\hUx)|$ approaches 1 as $\ell$ gets large, we arrive at the desired \rlb{oo2} and conclude that  the winding number is equal to $\nu_\omega$.
It is crucial to note that we already know that the winding number is well-defined (and independent of $\ell$) and are now using \rlb{oo} to determine its value.~\qedm

\medskip
Since the winding number is necessarily an integer, the above lemma leads to the following theorem.
\begin{theorem}[a necessary condition for a locally-unique gapped ground state]
Let $\omega$ be a locally-unique gapped ground state of a $\rm U(1)$ invariant Hamiltonian $\hH$.
We further assume that $\omega$ is invariant under translation by $p$.
Then the filling factor $\nu_\omega$ defined in \rlb{nu} must be an integer.
\end{theorem}

This is equivalent to the following no-go theorem for a locally-unique gapped ground state.

\begin{corollary}[generalized LSM theorem]\label{c:gLSM1D}
Consider a quantum spin system on the infinite chain with a $\rm U(1)$ invariant Hamiltonian $\hH$.
When the filling factor $\nu_\omega$ is not an integer, there can be no locally-unique gapped ground state that is invariant under translation by $p$.
\end{corollary}

It must be clear that these results are readily extended to models with site-dependent spin quantum numbers mentioned in footnote~\ref{fn:Sj}.
One only needs to replace the definition \rlb{nu} of filling factor by $\nu_\omega=\omega(\sum_{j=1}^p(\hSz_j+S_j))$.

The most important application of Corollary~\ref{c:gLSM1D}, which was discussed by Affleck and Lieb \cite{AL}, is the following condition for the antiferromagnetic Heisenberg chain \rlb{HAF} to have a unique gapped ground state.
The (global) uniqueness implies that the ground state is invariant under translation by 1, and also that $\omega(\hSz)=0$ and hence $\nu_\omega=S$.
Then Corollary~\ref{c:gLSM1D} reduces to the following.

\begin{corollary}[Affleck-Lieb theorem]\label{c:AL}
Let $S$ be a half-odd-integer.
Then it is impossible that the antiferromagnetic Heisenberg chain \rlb{HAF} has a unique gapped ground state.
\end{corollary}
As we have discussed at the beginning of this section, this proves the ``half'' of Haldane's famous conclusion that the antiferromagnetic Heisenberg chain has a unique gapped ground state if and only if $S$ is an integer \cite{Haldane1981,Haldane1983a,Haldane1983b,TasakiBook}.

%%%%%%%%%%%%%%%%
\section{Extended LSM theorem for translation-invariant spin chains with $\ZZ$ symmetry}
\label{s:ZZ}
%%%%%%%%%%%%%%%%

In the context of recent studies of topological phases of matter (see, e.g., \cite{ZengChenZhouWenBOOK}), it was realized that LSM-type no-go theorems are valid for quantum many-body systems that only possess certain discrete symmetry \cite{ChenGuWEn2011,PTAV,WPVZ,Fuji,PWJZ,Watanabe2018}.
In particular, as a part of their general classification theory, Chen, Gu, and Wen conjectured in sections~V.B.4 and  V.C of \cite{ChenGuWEn2011} that a translation-invariant quantum spin chain where the projective representation of the symmetry on each site is nontrivial cannot have a unique gapped ground state.
(See section~\ref{s:PR} below about projective representations.)
The statement for chains with time-reversal symmetry was proved by Watanabe, Po, Vishwanath, and Zaletel \cite{WPVZ} within the framework of matrix product states.
See also \cite{Prakash} and section~8.3.5 of \cite{TasakiBook} for general proofs for matrix product states.

In \cite{OT1}, Ogata and Tasaki confirmed the conjecture with full mathematical rigor for general translation-invariant quantum spin chains with $\ZZ$ or time-reversal symmetry.
The proof was an extension of the early work of Matsui \cite{Matsui1}, where the method based on the Cuntz algebra was developed.

Finally, in \cite{OTT}, Ogata, Tachikawa, and Tasaki gave a unified proof of the extended LSM theorem for general quantum spin chains.
The proof makes essential use of the topological index for quantum spin chains, now called the Ogata index, formulated recently by Ogata \cite{OgataZ2}.
In fact, given the basic definition and properties (which are nontrivial and not easy to prove) of the Ogata index, the proof in \cite{OTT} of the generalized LSM theorem is not difficult.

The Ogata index was introduced to discuss the classification of symmetry-protected topological (SPT) phases in quantum spin chains.
See, e.g., \cite{ZengChenZhouWenBOOK} and Part~II of \cite{TasakiBook} for background, and my online lecture  \cite{HalSPT} for a review of SPT phases and the Ogata index.

\medskip
The present section is intended to be a readable introduction to the proof of the extended LSM theorem by Ogata, Tachikawa, and Tasaki \cite{OTT}.
We start by discussing basic notions such as projective representations of a group and the corresponding index, and then explain the basic idea and the properties of the Ogata index without going into mathematical details.
See also the video of my seminar on the same subject \cite{HalIAMP}.

\subsection{Extended LSM theorem}
Let us start by presenting the main theorem.
We again follow the formulation of section~\ref{s:Gen} and study quantum spin systems on the infinite chain $\bbZ$.

For $\alpha=\xyz$, we denote by $\Gamma_\alpha$ the linear $*$-automorphism for the uniform $\pi$-rotation of spins about the $\alpha$-axis.
It is defined by
\eq
\Gamma_\alpha(\hS^{(\beta)}_j)=\begin{cases}
\hS^{(\beta)}_j,&\alpha=\beta;\\
-\hS^{(\beta)}_j,&\alpha\ne\beta,
\end{cases}
\en
for any $j\in\bbZ$ and $\alpha,\beta=\xyz$,
along with the basic properties of a linear $*$-automorphism stated in footnote~\ref{fn:*auto}.
As we shall see below in section~\ref{s:ZZPR}, $\Gamma_\mrx$, $\Gamma_\mry$, and $\Gamma_\mrz$ give a representation of the group $\ZZ$.

We say that a state $\rho$ is $\ZZ$-invariant, or, more precisely, invariant under $\ZZ$ transformation, if it holds that 
\eq
\rho(\Gamma_\alpha(\hA))=\rho(\hA),
\en
for any $\hA\in\Aloc$ and $\alpha=\xyz$.
Then the following necessary condition for a pure split state in a quantum spin chain was proved in \cite{OT1,OTT}.
\begin{theorem}\label{t:ZZ}
Assume that there exists a pure split state that is invariant under both $\ZZ$ transformation and translation by $p$.
Then it must be that $pS$ is an integer.\footnote{
The theorem is valid also for models with site-dependent spin quantum number mentioned in footnote~\ref{fn:Sj} if we replace $pS$ by $\sum_{j=1}^pS_j$.
}
\end{theorem}

Since  Theorem~\ref{t:puresplit} states that a locally-unique gapped ground state is necessarily a pure split state, we immediately get the following LSM-type no-go theorem.

\begin{corollary}[extended LSM theorem]\label{c:ZZ}
Consider a quantum spin system on the infinite chain.
When $pS$ is a half-odd-integer, there can be no locally-unique gapped ground state that is invariant under both $\ZZ$ transformation and translation by $p$.
\end{corollary}
Although we do not make any assumptions on the Hamiltonian, it is of course most meaningful to consider a Hamiltonian that is invariant under both $\ZZ$ transformation and translation.
The Hamiltonians of antiferromagnetic Heisenberg model \rlb{HAF} and the XXZ model \rlb{XXZ} with $H=0$ both satisfy this assumption with $p=1$.
An example of a Hamiltonian that is not U(1) invariant but $\ZZ$ invariant is
\eq
\hH_{\rm XYZ}=\sum_{j\in\bbZ}\bigl\{J_\mrx\,\hSx_j\hSx_{j+1}+J_\mry\,\hSy_j\hSy_{j+1}+J_\mrz\,\hSz_j\hSz_{j+1}\bigr\}.
\en

The same statements as Theorem~\ref{t:ZZ} and Corollary~\ref{c:ZZ} are valid if we replace the $\ZZ$ transformation with the time-reversal transformation.
It is described by the antlinear $*$-automorphism that is characterized by $\Gamma_{\rm tr}(\hSa_j)=-\hSa_j$ for any $j\in\bbZ$ and $\alpha=\xyz$.
See \cite{OT1,OTT} for details.

\subsection{Projective representations and the index of a finite group}
\label{s:PR}
In the rest of the section, we shall discuss the basic idea of the proof of Theorem~\ref{t:ZZ}.
We start with an elementary but general discussion about the classification of projective representations of a finite group.
Although we focus only on the group $\ZZ$ in the present article, we believe it useful to have a general picture in mind.

Let $G$ be a finite group with multiplication $\circ$, and let $\re$ denote its identity.
A representation (or, more precisely, a linear representation) of $G$ is given by a collection of unitary operators $\hU_g$ (on a certain Hilbert space) with $g\in G$ such that $\hU_\re=\hat{1}$ and $\hU_g\hU_h=\hU_{g\circ h}$ for any $g,h\in G$.
A projective representation  (or, more precisely, a linear projective representation) of $G$ is given by a collection of unitary operators $\hU_g$ (on a certain Hilbert space) with $g\in G$ such that $\hU_\re=\hat{1}$ and 
\eq
\hU_g\hU_h=e^{i\varphi(g,h)}\,\hU_{g\circ h},
\lb{PR}
\en
for any $g,h\in G$ with a phase factor $\varphi(g,h)\in[0,2\pi)$.
A representation is sometimes called a genuine representation as opposed to a projective representation.
Clearly a projective representation becomes a genuine representation when $\varphi(g,h)$ is always 0.

Since unitary operators satisfy the associativity $\hU_f(\hU_g\hU_h)=(\hU_f\hU_g)\hU_h$, the phase factor must satisfy
\eq
\varphi(g,h)+\varphi(f,g\circ h)=\varphi(f,g)+\varphi(f\circ g,h)\ ({\rm mod}\,2\pi),
\lb{2coc}
\en
for any $f,g,h\in G$.
This constraint \rlb{2coc} is known as the 2-cocycle condition, and $\varphi:G\times G\to[0,2\pi)$ satisfying the condition is called a 2-cocycle.
We denote by ${\rm Z}^2(G,\Uo)$ the set of all 2-cocycles, which naturally becomes an abelian group (where the group multiplication is simply the addition of phase factors).

The reader unfamiliar with the cohomology theory does not have to worry about these terminologies.
But we note in passing that the reader may be familiar with the following example of a 2-cocycle.
Consider the group $\bbZ_{10}=\{0,1,\ldots,9\}$, where the group multiplication $g\circ h$ is given by $g+h\ ({\rm mod}\,10)$.
It is then very useful to define a map $\sigma:\bbZ_{10}\times\bbZ_{10}\to\bbZ_{10}\times\bbZ_{10}$ as $\sigma(g,h)=(\varphi(g,h),g\circ h)$ with $\varphi:\bbZ_{10}\times\bbZ_{10}\to\bbZ_{10}$ satisfying the 2-cocycle condition $\varphi(g,h)+\varphi(f,g\circ h)=\varphi(f,g)+\varphi(f\circ g,h)$.
I learned this interpretation from Yuji Tachikawa, and an example of $\varphi$ from my first-grade teacher.\footnote{
The map $\sigma(g,h)$, which should better be written as $10\,\varphi(g,h)+g\circ h$, is usually called the addition, and the 2-cocycle $\varphi$ the carry.
See \cite{WikipediaCarry,Isaksen2002} and references therein for further details.
}

Suppose that there are two projective representations $\hU_g$ and $\hU'_g$ of $G$ associated with 2-cocycles $\varphi$ and $\varphi'$, respectively.
We say that two projective representations are equivalent when they differ only by a phase factor that depends on the group element, i.e., if there exists $\psi(g)\in[0,2\pi)$ such that $\hU'_g=e^{i\psi(g)}\hU_g$ for any $g\in G$.
In this case we see that the two 2-cocyles are related by
\eq
\varphi'(g,h)=\varphi(g,h)+\psi(g)+\psi(h)-\psi(g\circ h)\ ({\rm mod}\,2\pi).
\lb{equiv}
\en
This motivates us to define two 2-cocycles $\varphi$ and $\varphi'$ to be equivalent if \rlb{equiv} is valid for some $\psi_g$.
We then consider the quotient set (the set of equivalence classes) of ${\rm Z}^2(G,\Uo)$ with respect to this equivalence relation, and denote it as ${\rm H}^2(G,\Uo)={\rm Z}^2(G,\Uo)/\!\!\sim$.
The quotient set ${\rm H}^2(G,\Uo)$ is again regarded as an abelian group and called the second group cohomology of $G$.

In short, the second group cohomology ${\rm H}^2(G,\Uo)$ represents the set of equivalence classes of projective representations of $G$.
We denote by $\ind$ an element of  ${\rm H}^2(G,\Uo)$, and call it the index of the corresponding projective representation.

An important property of the index is additivity.
Suppose that there are two projective representations $\hU^{(1)}_g$ and $\hU^{(2)}_g$ on different Hilbert spaces characterized as $\hU^{(1)}_g\hU^{(1)}_h=e^{i\varphi_1(g,h)}\,\hU^{(1)}_{g\circ h}$ and $\hU^{(2)}_g\hU^{(2)}_h=e^{i\varphi_2(g,h)}\,\hU^{(2)}_{g\circ h}$.
We denote by $\ind_1$ and $\ind_2$ the corresponding indices, i.e., elements of ${\rm H}^2(G,\Uo)$.
Clearly the tensor product $\hU_g=\hU^{(1)}_g\otimes\hU^{(2)}_g$ also gives a projective representation satisfying 
\rlb{PR} with $\varphi(g,h)=\varphi_1(g,h)+\varphi_2(g,h)$.
This means that the index of the new projective representation is given by 
\eq
\ind=\ind_1+\ind_2.
\lb{indadd}
\en

%%%%%%%
\subsection{Projective representations and the index of the group $\ZZ$}
\label{s:ZZPR}
From now on we focus on the group $\ZZ$, which is relevant to us.
The group $\ZZ$, also known as the Klein group or the dihedral group $\rm D_2$, is an abelian group that consists of four elements $\re$, $\mrx$, $\mry$, and $\mrz$.
The multiplication rule is given by 
\eq
\mrx\circ\mrx=\mry\circ\mry=\mrz\circ\mrz=\re,\quad
\mrx\circ\mry=\mry\circ\mrx=\mrz,\quad
\mry\circ\mrz=\mrz\circ\mry=\mrx,\quad
\mrz\circ\mrx=\mrx\circ\mrz=\mry,
\lb{ZZR}
\en
as well as $g\circ\re=\re\circ g=g$ for any $g\in\ZZ$.
The elements $\mrx$, $\mry$, and $\mrz$ may be interpreted as the (spatial) $\pi$-rotation about the $\mrx$, $\mry$, and $\mrz$ axes, respectively.

It is well-known that the second group cohomology of $\ZZ$ is ${\rm H}^2(\ZZ,\Uo)=\bbZ_2=\{0,1\}$.
This means that there are exactly two equivalence classes of projective representations of $\ZZ$.
A projective representation with $\ind=0$ is equivalent to a genuine representation and is said to be trivial.
A projective representation with $\ind=1$ is said to be nontrivial.

Let us discuss important examples of projective representations of $\ZZ$, namely, those on a single quantum spin.
(See, e.g., chapter~2 of \cite{TasakiBook} for details.)
Consider a quantum spin with quantum number $S$, and let $\hSx$, $\hSy$, and $\hSz$ denote the spin operators.
We define $\hu^{(\re)}=\hat{1}$ and $\hug=\exp[-i\pi\hS^{(g)}]$ for $g=\xyz$.
It is easily checked that these unitary operators satisfy 
\eq
\hux\huy=\huz,\quad\huy\huz=\hux,\quad\huz\hux=\huy,
\en
recovering a part of the multiplication table \rlb{ZZR}.
 
When $S$ is an integer, these operators further satisfy $(\hug)^2=\hat{1}$ and $\hug\huh=\huh\hug$ for any $g,h\in\ZZ$.
This means that the collection $\{\hat{1},\hux,\huy,\huz\}$ faithfully recovers the multiplication rule \rlb{ZZR}, and hence gives a genuine representation.

When $S$ is a half-odd-integer, on the other hand, we see that $(\hug)^2=-\hat{1}$ and $\hug\huh=-\huh\hug$ for $g,h\in\{\xyz\}$ and $g\ne h$.
This means that $\{\hat{1},\hux,\huy,\huz\}$ gives a projective representation characterized by the 2-cocylcle $\varphi$ such that $\varphi(g,g)=\pi$ for $g\ne\re$, $\varphi(\mry,\mrx)=\varphi(\mrz,\mry)=\varphi(\mrx,\mrz)=\pi$, and $\varphi(g,h)=0$ otherwise.
Clearly, this $\varphi$ is not equivalent to the trivial 2-cocycle that is always zero.\footnote{
It suffices to note that \rlb{equiv} implies $\varphi'(g,h)-\varphi(g,h)=\varphi'(h,g)-\varphi(h,g)$ for an abelian group.
}
We thus have a nontrivial projective representation.

To summarize, the indices for the projective representations realized by a single quantum spin are given by
\eq
\ind=\begin{cases}
0,&S=1,2,\ldots;\\
1,&S=\frac{1}{2},\frac{3}{2},\ldots.
\end{cases}
\lb{indS}
\en

The above consideration can be readily generalized to a spin system on a finite lattice.
Consider a system of quantum spins on a finite interval $I$ of $\bbZ$, and let $\hug_j$ be the unitary operator at site $j\in I$ corresponding to $\hug$ with $g\in\ZZ=\{\re,\xyz\}$.
Obviously the tensor product $\hU^{(g)}_I=\bigotimes_{j\in I}\hug_j$ gives a projective representation of $\ZZ$ whose index is given simply by $\sum_{j\in I}\ind_j\in\{0,1\}$.

\subsection{Ogata index and the proof of Theorem~\ref{t:ZZ}}
Let $\omega$ denote a locally-unique gapped ground state (or, more generally, a pure split state) of a spin chain, and assume that $\omega$ is $\ZZ$ invariant.
Although the state $\omega$ on the whole chain $\bbZ$ does not change under the $\ZZ$ transformation, 
it may be the case that the same state restricted onto the half-infinite chain $\{j,j+1,\ldots\}$ exhibits nontrivial transformation properties.

\begin{figure}
\centerline{\epsfig{file=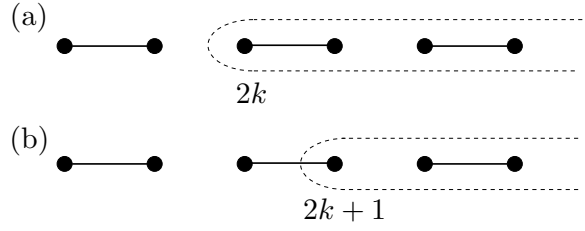,width=7.5truecm}}
\caption[dummy]{
A graphic representation of a part of the dimer state \rlb{dimerstate}.
A black dot represents a spin with $S=1/2$, and two dots connected by a line represents a spin-singlet.
(a)~The restriction of the state to the half-infinite chain $\{2k,2k+1,\ldots\}$ consists only of spin-singlets.  It is invariant under the $\ZZ$ transformation and should be characterized by the index $\Ind=0$.
(b)~The restriction of the state to the half-infinite chain  $\{2k+1,2k+2,\ldots\}$ contains an extra $S=1/2$ at site $2k+1$, and should be characterized by the index $\Ind=1$.
}
\label{f:dimer}
\end{figure}

To see that this is possible, consider a simple $S=1/2$ chain with the dimerized Hamiltonian
\eq
\hH_{\rm dimer}=\sum_{k\in\bbZ}\hat{\boldsymbol{S}}_{2k}\cdot\hat{\boldsymbol{S}}_{2k+1},
\lb{dimer}
\en
whose unique gapped ground state is a simple tensor product of spin-singlets, formally written as
\eq
\bigotimes_{k\in\bbZ}\frac{1}{\sqrt{2}}\Bigl(\ket{\!\!\up}_{2k}\ket{\!\!\downarrow}_{2k+1}
-\ket{\!\!\downarrow}_{2k}\ket{\!\!\up}_{2k+1}\Bigr).
\lb{dimerstate}
\en
Recall that a spin-singlet is invariant under the $\ZZ$ transformation, and hence gives a trivial genuine representation (in which $\hug=\hat{1}$ for any $g\in\ZZ$) with $\ind=0$.
If one restricts the ground state \rlb{dimer} to the half-infinite chain $\{2k,2k+1,\ldots\}$, one simply gets a simple tensor product of spin-singlets, which is $\ZZ$ invariant.
See Figure~\ref{f:dimer}~(a).
It is then natural to associate the ground state restricted onto $\{2k,2k+1,\ldots\}$ with the index $\Ind_{2k}=0$.
If one restricts the same ground state \rlb{dimer} on the half-infinite chain $\{2k+1,2k+2,\ldots\}$, on the other hand, one gets the mixture of $\ket{\sigma}_{2k+1}\otimes\{\bigotimes_{\ell>k}(\ket{\!\!\up}_{2\ell}\ket{\!\!\downarrow}_{2\ell+1}
-\ket{\!\!\downarrow}_{2\ell}\ket{\!\!\up}_{2\ell+1})/\sqrt{2}\}$ with $\sigma=\up,\downarrow$.
There is a single unpaired spin with $S=1/2$ at the edge.
This means that the ground state \rlb{dimer} restricted onto the half-infinite chain $\{2k+1,2k+2,\ldots\}$ transforms as a single $S=1/2$ spin under the $\ZZ$ transformation.
It is then natural to associate the restricted ground state with the index $\Ind_{2k}=1$.
See Figure~\ref{f:dimer}~(b).

\begin{figure}
\centerline{\epsfig{file=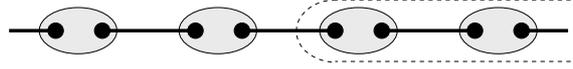,width=7.5truecm}}
\caption[dummy]{
A graphic representation of the exact ground state, known as the Valence-Bond Solid (VBS) state, of the AKLT Hamiltonian \rlb{AKLT}.
As in Figure~\ref{f:dimer}, a black dot represents a spin with $S=1/2$, and two dots connected by a line represents a spin-singlet.
Here two black dots surrounded by an oval represent the symmetrization of two $S=1/2$'s, which is equivalent to a state of spin with $S=1$.
If we examine the transformation property of the VBS state restricted onto a half-infinite chain, the situation is similar to that in Figure~\ref{f:dimer}~(b).
One observes an emergent $S=1/2$ degree of freedom at the edge, which should be characterized by the index $\Ind=1$.
See, e.g., Chapter~7 of \cite{TasakiBook} for details about the VBS state.
}
\label{f:VBS}
\end{figure}

A less trivial example is provided by the $S=1$ chain with the AKLT Hamiltonian
\eq
\hH_{\rm AKLT}=\sum_{j\in\bbZ}\Bigl\{\hat{\boldsymbol{S}}_{j}\cdot\hat{\boldsymbol{S}}_{j+1}+\frac{1}{3}(\hat{\boldsymbol{S}}_{j}\cdot\hat{\boldsymbol{S}}_{j+1})^2\Bigr\},
\lb{AKLT}
\en
which is proved to have a unique gapped ground state \cite{Matsui1997,AKLT1,AKLT2,TasakiBook}.
Moreover it is also known that the ground state restricted onto a half-infinite chain has an effective degrees of freedom of spin $S=1/2$ that emerges at the edge.
It is then natural to associate the restricted ground state with the index $\Ind_j=1$.
See Figure~\ref{f:VBS}.

Such indices for states restricted onto half-infinite chains were first defined by Pollmann, Turner, Berg, and Oshikawa for injective matrix product states \cite{PollmannTurnerBergOshikawa2010,PollmannTurnerBergOshikawa2012}.
The identification of the index was an essential step in the classification of symmetry-protected topological (SPT) phases.

\begin{figure}
\centerline{\epsfig{file=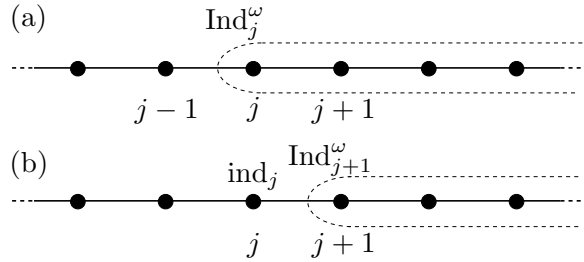,width=7.5truecm}}
\caption[dummy]{
A graphic interpretation of the additivity \rlb{Oadd}.
(a)~The Ogata index $\Ind_j^\omega$ characterizes the transformation property of the state $\omega$ restricted onto the half-infinite chain $\{j,j+1,\ldots\}$.
(b)~The same transformation property can be expressed by the sum of the index $\ind_j$ of the single spin at site $j$ and the Ogata index $\Ind^\omega_{j+1}$ for the half-infinite chain $\{j+1,j+2,\ldots\}$.
}
\label{f:add}
\end{figure}

In 2019, Ogata extended the index to general locally-unique gapped ground states, or, more precisely, to general pure split states with certain symmetry of a spin chain \cite{OgataZ2}.
Although Ogata's theory applies to a general symmetry group, we here concentrate on the case with $\ZZ$ symmetry.

Let $\omega$ be a pure split state that is invariant under the $\ZZ$ transformation.
By using machinery in the operator algebraic formulation of quantum spin systems, one can construct a Hilbert space $\calH_j$ and unitary operators $\hU^{(\mrx)}_j$, $\hU^{(\mry)}_j$, and $\hU^{(\mrz)}_j$ that represent the transformation property of the state $\omega$ restricted onto the half-infinite chain $\{j,j+1,\ldots\}$.
The unitary operators form a projective representation of $\ZZ$, whose index is the Ogata index $\Ind^\omega_j\in\bbZ_2$.
See Figure~\ref{f:add}~(a).
See section~8.3.6 of \cite{TasakiBook} for some more details, and \cite{OgataZ2,OgataCDM} for full details.
Ogata showed that $\Ind^\omega_j$ coincides with the index of Pollmann, Turner, Berg, and Oshikawa for injective matrix product states.
Moreover, Ogata proved that the index  $\Ind^\omega_j$ is invariant under smooth modifications of unique gapped ground states, thus essentially completing the general classification theory of SPT phases in quantum spin chains \cite{OgataZ2,OgataCDM}.

An essential (and nontrivial) property of the Ogata index is additivity.
Recall tha the index $\Ind^\omega_j$ characterizes the transformation property of the state restricted onto the half-infinite chain $\{j,j+1,\ldots\}$.
Since $\{j,j+1,\ldots\}$ is decomposed into a single site $j$ and the half-infinite chain $\{j+1,j+2,\ldots\}$, it is expected that the Ogata index satisfies the additivity as in \rlb{indadd}, i.e.,
\eq
\Ind^\omega_j=\ind_j+\Ind^\omega_{j+1},
\lb{Oadd}
\en
where $\ind_j$ is the index for a single spin given by \rlb{indS}.
See Figure~\ref{f:add}.
The identity \rlb{Oadd} was proved by Ogata.  See \cite{OTT}.

We are now ready to prove the main theorem of this section.

\medskip
\noindent
{\em Proof of Theorem~\ref{t:ZZ}:}\/
Let $\omega$ be a $\ZZ$ invariant pure split state, and let $\Ind^\omega_j$ denote the Ogata index.
We further assume that the state is invariant under translation by $p$, in which case we have
\eq
\Ind^\omega_1=\Ind^\omega_{1+p}.
\en
This equality and \rlb{Oadd} then implies
\eq
\sum_{j=1}^p\ind_j=\Ind^\omega_1-\Ind^\omega_{1+p}=0.
\en
Recalling \rlb{indS} and the additivity \rlb{indadd}, we see that $\sum_{j=1}^p\ind_j=0$ implies $pS\in\bbZ$.~\qedm

%%%%%%%%%%%%%%%%%%%
\section{LSM-type theorems for quantum spin systems on the infinite cylinder}
\label{s:2D}
%%%%%%%%%%%%%%%%%%%%
It goes without saying that to prove LSM-type no-go theorems for higher-dimensional quantum many-body systems is extremely important.
In fact, Lieb, Schultz, and Mattis discussed in their original paper \cite{LSM} that their method can be readily extended to certain higher-dimensional models with strong anisotropy.
Such an extension was further discussed by Affleck \cite{Affleck1988}.

In the present section, we follow this strategy and briefly discuss generalized and extended LSM theorems for quantum spin systems defined on the infinite cylinder.
We also observe in section~\ref{s:spiral} that, by imposing the spiral (or tilted) boundary conditions, one gets the desired filling factor in the conditions of the theorems.

We note that these theorems apply only to systems that are essentially one-dimensional.
See section~\ref{s:discussion} for a brief survey of full-fledged LSM theorems for higher-dimensional models.

\subsection{Quantum spin systems on the infinite cylinder}
\begin{figure}
\centerline{\epsfig{file=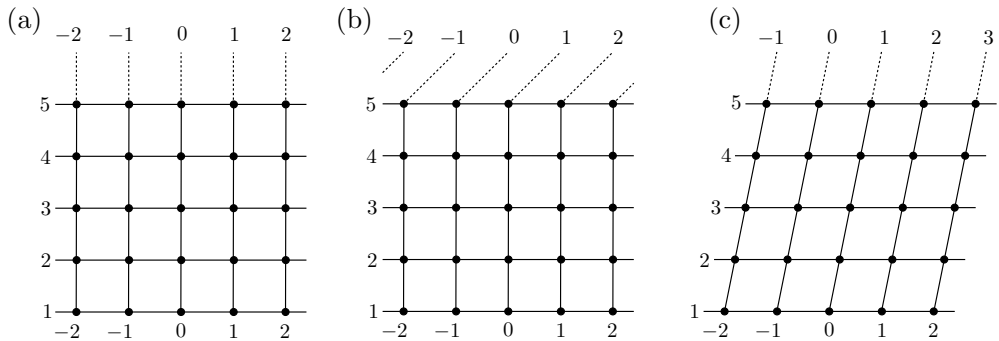,width=13truecm}}
\caption[dummy]{
Boundary conditions for the anisotropic two-dimensional lattice $\LaL$ with $L=5$.
(a)~In the standard periodic boundary conditions, one identifies $(j_1,L+1)$ with $(j_1,1)$.
(b)~In the spiral boundary condition with $p_1=1$, one identifies $(j_1,L+1)$ with $(j_1+1,1)$.
(c)~The same spiral boundary conditions may be represented as the tilted periodic boundary conditions \cite{YaoOshikawa}.
}
\label{f:BC}
\end{figure}

Take an anisotropic two-dimensional lattice $\LaL=\bbZ\times\{1,\ldots,L\}$, which is infinite in one direction and finite in the other direction.
A site in $\LaL$ is denoted as $\pjj$ with $j_1\in\bbZ$ and $j_2=1,\ldots,L$.
It is standard to impose the periodic boundary conditions to identify $(j_1,L+1)$ with $(j_1,1)$, and regard $\LaL$ as a cylinder.
See Figure~\ref{f:BC}~(a).
One can also employ the open boundary conditions and regard $\LaL$ as a strip.

We consider a quantum spin system with spin quantum number $S$ on $\LaL$, and denote by $\hSx_{\jj}$, $\hSy_{\jj}$, and $\hSz_{\jj}$ the spin operators at site $\pjj$.
The Hamiltonian is again formally written as an infinite sum
\eq
\hH=\sum_{\pjj\in\LaL}\hh_{\jj},
\lb{H2d}
\en
where the local Hamiltonian $\hh_{\jj}$ is assumed to be short-ranged and uniformly bounded as in section~\ref{s:setting1D}.

In this setting, one can define the notions of states, ground states, and a (locally-)unique gapped ground state exactly as in section~\ref{s:setting1D}.
We shall again look for no-go theorems for a locally-unique gapped ground state.

\subsection{Generalized LSM theorem for U(1) invariant models}
\label{s:GLSM2D}
Let us assume that the local Hamiltonian $\hh_{\jj}$ with any $\pjj$ is U(1) invariant as in \rlb{U1}.
For any $x\in\bbR$ and $\ell>0$, we follow \rlb{Uxl} and define the local twist operator by
\eq
\hUx=\exp\Bigl[-i\sum_{j_1\in\bbZ}\sum_{j_2=1}^L\theta_{j_1}(\hSz_{\jj}+S)\Bigr],
\lb{Uxl2d}
\en
where $\theta_j$ is the same as \rlb{theta}.
Note that the spin-rotation angle stays constant in the 2-direction and varies only in the 1-direction.
Then by repeating the proof of Lemma~\ref{l:var}, we get
\eq
0\le\omega(\hUx^\dagger\,[\hH,\hUx])\le\frac{C L}{\ell},
\lb{var2}
\en
for any ground state $\omega$, where $C$ is a constant.
The bound \rlb{var2} is the same as \rlb{var} except that the factor $C$ is replaced by $CL$.
Clearly the right-hand side of \rlb{var2} is not small when $L$ and $\ell$ are comparable.
But, by making use of the cylindrical geometry of the lattice, one can simply regard $CL$ as a new constant, and take $\ell$ to be much larger than $CL$.
Then everything is the same as in the one-dimensional case.

Let us focus on ground states that are invariant under translation by $p$ in the 1-direction.
Then the relevant ``filling factor'' that corresponds to \rlb{nu} is
\eq
\nu^L_\omega=\omega\Bigl(\sum_{j_1=1}^p\sum_{j_2=1}^L(\hSz_{\jj}+S)\Bigr).
\lb{nu2d}
\en
By exactly repeating the logic in section~\ref{s:U1LSM}, we arrive at the following generalized LSM theorem that corresponds to Corollary~\ref{c:gLSM1D}.

\begin{theorem}[generalized LSM theorem on the cylinder]\label{t:gLSM2D}
Consider a quantum spin system on the infinite cylinder $\LaL=\bbZ\times\{1,\ldots,L\}$ with the periodic or open boundary conditions with a $\rm U(1)$ invariant Hamiltonian $\hH$.
When $\nu^L_\omega$ is not an integer, there can be no locally-unique gapped ground state that is invariant under translation by $p$ in the 1-direction.
\end{theorem}

We should note that the ``filling factor'' $\nu^L_\omega$ defined in \rlb{nu2d} is indeed not a quantity that one would expect in a genuine two-dimensional LSM theorem.
In two dimensions, it is natural to focus on ground states that are invariant under translation by $p_1$ in the 1-direction and translation by $p_2$ in the 2-direction.
Then one defines the filling factor as
\eq
\nu_\omega=\omega\Bigl(\sum_{j_1=1}^{p_1}\sum_{j_2=1}^{p_2}(\hSz_{\jj}+S)\Bigr),
\lb{nu2d2}
\en
which is the total magnetization (or the particle number) in the unit cell with $p_1p_2$ sites.
It is expected that any translation-invariant locally-unique gaped ground state $\omega$ has an integral filling factor $\nu_\omega$.
On the contrary to \rlb{nu2d2}, the ``filling factor'' $\nu^L_\omega$ defined in \rlb{nu2d} represents the total magnetization in the region with $pL$ sites, which could be large if the strip is wide (and close to two-dimension).
Note, in particular, that the statement corresponding to the Affleck-Lieb theorem (Corollary~\ref{c:AL}) requires $SL$ to be a half-odd-integer.
The condition may be satisfied when $L$ is odd, but never be satisfied when $L$ is even.
But the nature of the ground state is likely independent of the parity of $L$ when $L$ is sufficiently large, provided that the ground state does not exhibit antiferromagnetic long-range order.

In section~\ref{s:spiral} we shall see that the filling factor \rlb{nu2d2} appears in the theory if we take different boundary conditions, namely, the spiral (or tilted) boundary conditions.

\subsection{Extended LSM theorem for $\ZZ$ invariant models}
The extended LSM theorem for $\ZZ$ invariant locally-unique gapped ground states that we discussed in section~\ref{s:ZZ} can also be extended to the present geometry.

To see this, note that the lattice $\LaL$ may be identified with the infinite chain $\bbZ$ by a one-to-one map as $(j_1,j_2)\mapsto L j_i+j_2$.
We can thus regard any quantum spin system on $\LaL$ as a quantum spin chain.
Note that any short-ranged Hamiltonian on $\LaL$ is mapped to a (somewhat complicated) short-ranged Hamiltonian on $\bbZ$.
This means that Theorem~\ref{t:puresplit}, which is essential for the use of the Ogata index, and all the general results about the Ogata index are still valid in the present class of models.
Finally noting that the translation by $p$ in the 1-direction for $\LaL$ corresponds to the translation by $pL$ on $\bbZ$, we get the following extended LSM theorem.

\begin{theorem}[extended LSM theorem on the cylinder]\label{c:ZZ2}
Consider a quantum spin system on the infinite cylinder $\LaL=\bbZ\times\{1,\ldots,L\}$ with the periodic or open boundary conditions.
When $pLS$ is a half-odd-integer, there can be no locally-unique gapped ground state that is invariant under both $\ZZ$ transformation and translation by $p$ in the 1-direction.
\end{theorem}

%%%%%%%%
\subsection{LSM-type theorems for models with spiral boundary conditions}
\label{s:spiral}

Here we follow Yao and Oshikawa \cite{YaoOshikawa}, and discuss different boundary conditions called the spiral (or the tilted) boundary conditions.
The same boundary conditions are used also in \cite{NakamuraMasudaNishimoto}.
Although we here concentrate on the (highly anisotropic) two-dimensional systems, one may devise analogous boundary conditions for systems with higher dimensions.

Fix the periods $p_1$ and $p_2$.
The most basic choice is $p_1=p_2=1$.
We take $L$ to be an integer multiple of $p_2$ and consider the same anisotropic lattice $\LaL=\bbZ\times\{1,\ldots,L\}$.
We then impose the spiral boundary conditions (or tilted periodic boundary conditions) by identifying $(j_1,L+1)$ with $(j_1+p_1,1)$.
See Figure~\ref{f:BC}~(b), (c).

\begin{figure}
\centerline{\epsfig{file=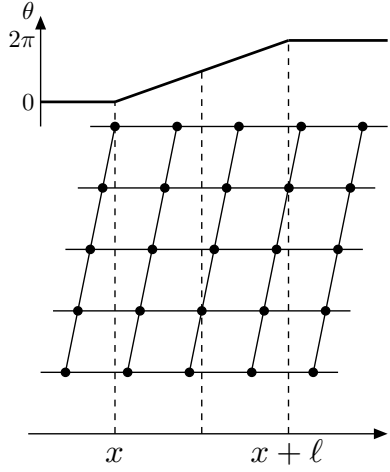,width=5truecm}}
\caption[dummy]{
The rotation angle $\theta_{\jj}$ of \rlb{theta2} is determined by the horizontal coordinate in this figure.
}
\label{f:spiraltheta}
\end{figure}

Let $\calT_{k_1,k_2}$ be the linear $*$-automorphism for the translation by $k_1$ in the 1-direction and $k_2$ in the 2-direction, which is  defined by $\calT_{k_1,k_2}(\hSa_{\jj})=\hSa_{j_1+k_1,j_2+k_2}$.
Here we take into account the spiral boundary conditions when considering translation in the 2-direction.

We agin consider a short-ranged and uniformly bounded U(1) invariant Hamiltonian \rlb{H2d}.
For any $x\in\bbR$ and $\ell>0$, we introduce new local twist operator as in \cite{NakamuraMasudaNishimoto}
\eq
\hUx=\exp\Bigl[-i\sum_{j_1\in\bbZ}\sum_{j_2=1}^L\theta_{\jj}(\hSz_{\jj}+S)\Bigr],
\lb{Uxl2d2}
\en
where the rotation angle $\theta_{\jj}$ is chosen to be compatible with the spiral boundary conditions as
\eq
\theta_{\jj}=\begin{cases}
0,&j_1< x-\frac{p_1j_2}{L};\\
2\pi(j-x+\frac{p_1j_2}{L})/\ell,&x-\frac{p_1j_2}{L}\le j_1\le x-\frac{p_1j_2}{L}+\ell;\\
2\pi,&j_1>x-\frac{p_1j_2}{L}+\ell.
\end{cases}
\lb{theta2}
\en
See Figure~\ref{f:spiraltheta}.
Again by using the same estimate as in the proof of Lemma~\ref{l:var}, we get
\eq
0\le\omega(\hUx^\dagger\,[\hH,\hUx])\le\frac{C' L}{\ell},
\lb{var3}
\en
for any ground state $\omega$, where $C'$ is a constant.

Let us now assume that the ground state $\omega$ is invariant under translation by $p_2$ in the 2-direction.
Recall that $L$ is an integer multiple of $p_2$, and $\calT_{0,L}=\calT_{p_1,0}$ because of the spiral boundary conditions.
We then see that $\omega$ is also invariant under translation by $p_1$ in the 1-direction.
With the periodicity in mind, it is natural to consider the filling factor defined in \rlb{nu2d2}, the total particle number in the unit cell with $p_1p_2$ sites.

From \rlb{Uxl2d2} and \rlb{theta2}, one finds (by inspection) that
\eq
\calT_{0,-p_2}(\hUx)=\hU_{x+p_1p_2/L,\ell}.
\en
Then from the translation invariance of the ground state $\omega$, we get the key relation
\eq
\omega(\hU_{0,\ell})=\omega(\hU_{p_1p_2/L,\ell}),
\lb{p1p2}
\en
which plays the role of \rlb{oUoU} for one-dimension.
The rest is, again, the same as before.
We assume that $\omega$ is a locally-unique gapped ground state, and use the variational estimate \rlb{var3} to show that $\omega(\hUx)\ne0$ for sufficiently large $\ell$.
Then the invariance \rlb{p1p2} implies that the winding number is well-defined.
One finally shows, again as in the one-dimensional case, that the winding number is nothing but the filling factor \rlb{nu2d2}.

This leads us to the following generalized LSM theorem in which the condition is written in terms of the desirable filling factor.

\begin{theorem}[generalized LSM theorem for the spiral boundary conditions]\label{t:gLSM2Ds}
Consider a quantum spin system on the infinite cylinder $\LaL=\bbZ\times\{1,\ldots,L\}$ with the spiral boundary conditions (corresponding to the periods $p_1$ and $p_2$) with a $\rm U(1)$ invariant Hamiltonian $\hH$.
When $\nu_\omega$ of \rlb{nu2d2} is not an integer,  there can be no locally-unique gapped ground state that is is invariant under translation by $p_1$ in the 1-direction and that by $p_2$ in the 2-direction.\footnote{\label{fn:12}The translation invariance in the 1-direction automatically follows from that in the 2-direction.}
\end{theorem}

Likewise, the statement corresponding to Corollary~\ref{c:AL} now only involves the spin quantum number $S$, rather than $LS$.

\begin{corollary}[Affleck-Lieb theorem for the spiral boundary conditions]\label{c:AL2}
Let $S$ be a half-odd-integer.
Then it is impossible that the antiferromagnetic Heisenberg model (with uniform nearest neighbor interaction) on the infinite cylinder $\LaL=\bbZ\times\{1,\ldots,L\}$ with the spiral boundary conditions for $p_1=p_2=1$ has a unique gapped ground state.
\end{corollary}

The validity of these theorems is not surprising if one notes that the cylindrical lattice $\LaL$ with the spiral boundary conditions can be regarded as consisting of $p_1$ infinite chains that spirally wrap around the infinite cylinder.
Then our problem reduces to that of a quantum spin chain that contains interactions of range $L$ as well as short-ranged interactions.
One also finds that the twist operator  \rlb{Uxl2d2} is simply the standard one-dimensional twist operator \rlb{Uxl}, especially when $p_1=1$.

From this mapping one immediately gets the following theorem for models with discrete symmetry.

\begin{theorem}\label{c:ZZsp}
Consider a quantum spin system on the infinite cylinder $\LaL=\bbZ\times\{1,\ldots,L\}$ with the spiral boundary conditions (corresponding to the periods $p_1$ and $p_2$).
When $p_1p_2S$ is a half-odd-integer, there can be no locally-unique gapped ground state that is invariant under $\ZZ$ transformation and translation by $p_1$ in the 1-direction and that by $p_2$ in the 2-direction.
\end{theorem}

%%%%%%%%%%
\section{Discussion}
\label{s:discussion}
In this review article, we discussed the generalized LSM theorem for U(1) invariant spin chains and the extended LSM theorem for $\ZZ$ invariant spin chains.
Both the theorems are proved by examining characteristic necessary conditions for the existence of a translation-invariant locally-unique gapped ground state.
The necessary conditions are expressed in terms of topological indices that characterize a locally-unique gapped ground state with necessary symmetry.
We hope that, in the case of U(1) symmetric chains, this rearrangement of the original strategy by Lieb, Schultz, and Mattis is of interest and enlightening.
We also noted that these theorems rule out locally-unique gapped ground states, not merely unique gapped ground states.

Although we only treated quantum spin systems in the present article, LSM-type theorems for quantum particle systems on lattices are discussed in the literature.
Quantum particle systems with number conservation law (which are natural as models in condensed matter physics or ultracold atom physics) have built-in U(1) symmetry, which can be used to define twist operators (as was originally done by Bloch \cite{Bohm,TadaKoma,Watanabe2019}).
Such a generalization of the LSM theorem was first discussed by Yamanaka, Oshikawa, and Affleck \cite{YOA}.
See also \cite{TasakiLSM} where quantum spin systems and lattice electron systems are treated in a unified manner.

Let us finally make some comments on LSM-type theorems for higher-dimensional systems.

As is clear from the proof, the generalized and extended LSM theorems for systems on the infinite cylinder that we discussed in section~\ref{s:2D} are essentially one-dimensional theorems, which apply only to highly anisotropic systems.
This point is most clearly seen from the fact that the same arguments do not produce any meaningful results for a system on the infinite two-dimensional lattice $\bbZ^2$ or the finite square lattice.

In 1999, based on the flux-insertion argument, Oshikawa proposed an intrinsically higher-dimensional version of the LSM theorem for U(1) invariant systems \cite{O}.
By using a different argument, Hastings proved the LSM theorem for a class of quantum spin systems that includes the Heisenberg antiferromagnet on a finite higher-dimensional lattice \cite{H1}.  (See also \cite{H2}.)
Hastings' proof was refined and made rigorous by Nachtergaele and Sims \cite{NS}. 
Later, as an application of their index theorem for U(1) invariant quantum many-body systems, Bachmann, Bols, De Roeck, and Fraas proved a generalized LSM theorem for a larger class of higher-dimensional systems \cite{BachmannBolsDeRoecFraas2019}.
The theorem in \cite{H1,NS} provides an explicit upper bound for the first excitation energy above the unique ground state in a finite system.
This is analogous to the original theorem by Lieb, Schultz, and Mattis \cite{LSM}.
Bachmann, Bols, De Roeck, and Fraas, on the other hand, directly prove a no-go theorem from a necessary condition for the existence of a unique-gapped ground state \cite{BachmannBolsDeRoecFraas2019}.
Their proof may be regarded as a rigorous version of Oshikawa's argument \cite{O}, although the connection is not explicit.

We should note that these higher-dimensional LSM theorems \cite{H1,NS,BachmannBolsDeRoecFraas2019} apply to arbitrary finite systems but not to the infinite system.
This is closely related to the fact that the quantization conditions in these theorems may not yet be optimal.
To be precise, the theorem of Hastings and Nachtergaele-Sims shows that the spin $S$ Heisenberg antiferromagnet model on the $L_1\times L_2$ two-dimensional lattice has a low-energy excited state above the ground state provided that $L_1$ is even and $L_2 S$ is a half-odd-integer.
Likewise, the theorem of Bachmann, Bols, De Roeck, and Fraas shows that the existence of a  unique gapped ground state requires $L_2\{\omega(\hSz_{j_1,j_2})+S\}$ to be an integer.
As we discussed in section~\ref{s:GLSM2D}, these may not be the optimal conditions when the model has translation invariance in both the 1 and the 2-directions.
In fact, these theorems make use only of the translation invariance in the 1-direction.

\begin{figure}
\centerline{\epsfig{file=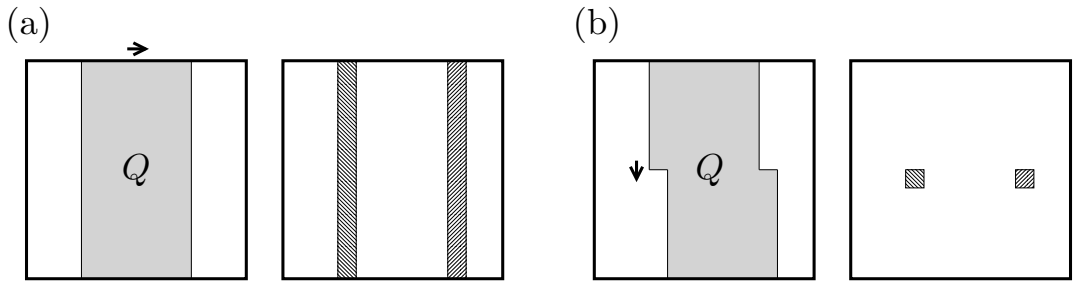,width=14truecm}}
\caption[dummy]{
These figures are for the reader familiar with the work of Bachmann, Bols, De Roeck, and Fraas \cite{BachmannBolsDeRoecFraas2019}.
(a)~In the original setting in Example~2 of \cite{BachmannBolsDeRoecFraas2019}, one works on a system with the periodic boundary conditions, and lets $\hQ=\hQ_\Gamma$ be the total U(1) charge in the gray region.
If we denote by $\hU$ the unitary for translation (by unit lattice spacing) in the 1-direction, then $\hU^\dagger\hQ\hU-\hQ$ is the difference in the total charges in the two strip-shaped regions.
In a unique gapped ground state, one obtains the quantization condition $L_2\{\omega(\hSz)+S\}\in\bbZ$.
(b)~We here consider a system with the spiral boundary conditions in the 2-direction, and let $\hQ$ be the total U(1) charge in the gray region.
If we denote by $\hU$ the unitary for the translation in the 2-direction, then $\hU^\dagger\hQ\hU-\hQ$ is the difference in the charges in the two plaquettes.
In a unique gapped ground state, the index theorem of \cite{BachmannBolsDeRoecFraas2019} yileds the desirable quantization condition $\omega(\hSz)+S\in\bbZ$.
}
\label{f:IndexSpiral}
\end{figure}

We remark that the situation about the quantization condition may be improved if one uses the spiral boundary condition of Yao and Oshikawa \cite{YaoOshikawa,NakamuraMasudaNishimoto} that we discussed in section~\ref{s:spiral}.
Take the $L_1\times L_2$ square lattice and impose the periodic boundary conditions in the 1-direction and the spiral boundary conditions in the 2-direction\footnote{
Denote the lattice sites as $(j_1,j_2)$ with $j_1=1,\ldots,L_1$ and $j_2=1,\ldots,L_2$.
We identify $(L_1+1,j_2)$ with $(1,j_2)$, and $(j_1,L+1)$ with $(j_1+1,1)$.
Note that the lattice becomes bipartite when $L_1$ is even and $L_2$ is odd.
One can also devise analogous boundary conditions in higher dimensions.
}.
Then the general index theorem of Bachmann, Bols, De Roeck, and Fraas implies that a quantum spin system invariant under the U(1) transformation and the translation in the 2-direction can have a unique gapped ground state only when $\omega(\hSz)+S$ is an integer \cite{Bachmann}.
See Figure~\ref{f:IndexSpiral}.
This is the quantization condition expected from heuristic arguments \cite{O}.
One should of course note that this is a consequence of the very special boundary conditions.
In fact, it seems to be still very difficult to prove the corresponding theorem for the infinite system.

Extended LSM theorems for models with discrete symmetry in higher-dimensional systems are also expected to be valid \cite{PTAV,WPVZ,PWJZ,Watanabe2018,YaoOshikawa,YaoOshikawa2}, but there are no rigorous results at the time of writing.

%%%%%%%%%%%%%%%%%
\appendix
\section{Proof of Theorem~\ref{t:puresplit}}
\label{s:Yoshiko}
In this Appendix, we prove Theorem~\ref{t:LUGGS}, which provides an essential characterization of a locally-unique gapped ground state.
In one dimension this theorem allows us to use Matsui's result\footnote{
For this purpose, the improvement in \cite{Matsui2013} of the earlier result in \cite{Matsui2010} is essential.
} in \cite{Matsui2013} to prove Theorem~\ref{t:puresplit}, which played an essential role in section~\ref{s:ZZ}.
Although we discuss applications only in one dimension, Theorem~\ref{t:LUGGS} itself is not limited to one dimension.
We here formulate and prove it for a general $d$-dimensional quantum spin system.
Unlike in the main text, we here assume basic knowledge on the operator algebraic formulation of quantum spin systems found, e.g., in \cite{BR1,BR2}.
The material in this appendix is due to Yoshiko Ogata.

Consider a quantum spin system on the infinite $d$-dimensional lattice $\bbZ^d$ defined by associating each site $p\in\bbZ^d$ with a quantum spin with the spin quantum number $S$ described the spin operator $(\hSx_p,\hSy_p,\hSz_p)$.
We again denote by $\Aloc$ the set of all polynomials of spin operators and by $\OA$ its completion with respect to the operator norm.
The dynamics of the spin system is determined by the formal Hamiltonian
\eq
\hH=\sum_{p\in\bbZ^d}\hh_p,
\en
or, more precisely, by the collection of local Hamiltonians $\hh_p$ with $p\in\bbZ^d$.
We again assume that the local Hamiltonians are short-ranged and uniformly bounded, i.e., $\hh_p$ depends only on spin operators $\hSa_q$ with $q$ such that $|p-q|\le r_0$ and satisfies $\snorm{\hh_p}\le h_0$, with constants $r_0$ and $h_0$.
Then it is known that there exists a one-parameter family of linear $*$-automorphisms on $\OA$, which we denote as $\tau^t_{\hH}$, that describes the time evolution of operators by $t\in\bbR$.
We denote by $\delta_{\hH}$ the generator of  $\tau^t_{\hH}$.
For $\hA\in\Aloc$, we have
\eq
\delta_{\hH}(\hA)=i\Bigl[\sum_{p\in\tilde{\Lambda}_L}\hh_p,\hA\Bigr],
\en
for sufficiently large $L$, 
where $\tilde{\Lambda}_L=\{-L,-L+1,\ldots,L\}^d$.

Let us repeat the definitions of ground states and a locally-unique gapped ground state.
\begin{definition}[ground states]\label{def:gsg}
A state $\omega$ on $\OA$ is a ground state if 
\eq
-i\,\omega(\hV^\dagger\delta_{\hH}(\hV))\ge0,
\en
for any $\hV\in\Aloc$.
\end{definition}
\begin{definition}[locally-unique gapped ground state]
A ground state\footnote{
This assumption is in fact redundant since one can prove that a state $\omega$ satisfying the condition \rlb{gamma} is automatically a ground state.
But we keep this assumption for simplicity.
} $\omega$ is a locally-unique gapped ground state if there exists $\gamma>0$ such that
\eq
-i\,\omega(\hV^\dagger\delta_{\hH}(\hV))\ge\gamma\,\omega(\hV^\dagger\hV),
\lb{gamma}
\en
for any $\hV\in\Aloc$ with $\omega(\hV)=0$.
\end{definition}

We note that Theorem~\ref{t:finiteGS} about finite volume ground states readily extend to $d$-dimensional systems.
It is also true in general that a locally-unique gapped ground state is not necessarily a unique gapped ground state.
Kitaev's toric code model on $\bbZ^2$ \cite{Kitaev} (see section~8.4 of \cite{TasakiBook} for an introduction) provides a nontrivial example of a locally-unique gapped ground state that is not a unique ground state.\footnote{
\label{fn:toric}
The toric code model on a finite square lattice with open boundary conditions has a unique frustration-free gapped ground state.
The infinite volume limit of these ground states defines a frustration-free ground state of the toric code model on $\bbZ^2$.
From the extension of Theorem~\ref{t:finiteGS}, we see that this limiting ground state is locally-unique and gapped.
However, it was proved in \cite{ChaNaaijkensNachtergaele} that the toric code model on $\bbZ^2$ has exactly four ground states in the sense of Definition~\ref{def:gsg}.
Three other ground states, which are not frustration-free, are characterized by the presence of an anyon.
}

The main result in the present appendix is the following.
\begin{theorem}\label{t:LUGGS}
Let $\omega$ be a locally-unique gapped ground state.
In the GNS representation corresponding to $\omega$, the GNS Hamiltonian has a nondegenerate ground state accompanied by a nonzero gap.
Furthermore, $\omega$ is a pure state.
\end{theorem}
If we restrict ourselves to one-dimensional systems, the theorem allows us to use the result of Matsui, stated as Corollary~3.2 of \cite{Matsui2013}, to conclude that $\omega$ satisfies the split property.
This proves Theorem~\ref{t:puresplit}.

\medskip
\noindent{\em Proof of Theorem~\ref{t:LUGGS}:}\/
Let $(\Hiw,\piw,\Ow)$ be the GNS triple corresponding to $\omega$, and let $\bktw{\cdot,\cdot}$ denote the inner product on $\Hiw$.
Recall that $\omega(\hA)=\bktw{\,\Ow,\piw(\hA)\,\Ow}$ for any $\hA\in\OA$.
It is known that there exists a unique nonnegative operator $\Hw$ on $\Hiw$, which we call the GNS Hamiltonian, that reproduces the time-evolution as
\eq
e^{it\Hw}\piw(\hA)\,\Ow=\piw(\tau^t_{\hH}(\hA))\,\Ow,
\en
for any $\hA\in\OA$ and $t\in\bbR$.
In terms of the generator, the relation reads
\eq
\Hw\,\piw(\hA)\,\Ow=-i\,\piw(\delta_{\hH}(\hA))\,\Ow,
\lb{A1}
\en
for any $\hA\in\Aloc$.
Note that $\Ow$ is an eigenvector of $\Hw$ with eigenvalue 0.

Let $P_\Omega$ be the orthogonal projection onto the space $\bbC\,\Ow$, and let $Q_\Omega=\id-P_\Omega$.
We then have for any $\hA\in\OA$ that
\eq
Q_\Omega\,\piw(\hA)\,\Ow=\piw(\hA)\,\Ow-\bktw{\Ow,\piw(\hA)\,\Ow}\,\Ow=
\piw(\hA-\omega(\hA)\,\hat{1})\,\Ow.
\lb{A2}
\en
For $\hA\in\Aloc$, let $\hV=\hA-\omega(\hA)\,\hat{1}$.
Since $\omega(\hV)=0$ we see from \rlb{gamma} that
\eq
-i\,\omega\bigl((\hA-\omega(\hA)\,\hat{1})^\dagger\,\delta_{\hH}(\hA-\omega(\hA)\,\hat{1})\bigr)\ge
\gamma\,\omega\bigl((\hA-\omega(\hA)\,\hat{1})^\dagger\,(\hA-\omega(\hA)\,\hat{1})\bigr).
\en
By using \rlb{A1} and \rlb{A2}, this is rewritten as
\eq
\bktw{Q_\Omega\,\piw(\hA)\,\Ow,\Hw\,Q_\Omega\,\piw(\hA)\,\Ow}\ge\gamma\,\snorm{Q_\Omega\,\piw(\hA)\,\Ow}^2,
\en
which, by exchanging $\Hw$ and $Q_\Omega$, reads
\eq
\bktw{Q_\Omega\,\piw(\hA)\,\Ow,Q_\Omega\,\Hw\,\piw(\hA)\,\Ow}\ge\gamma\,\snorm{Q_\Omega\,\piw(\hA)\,\Ow}^2.
\lb{A3}
\en

We are ready to prove that 0 is a nondegenerate eigenvalue of $\Hw$.
Assume that $\ker\Hw\ne\bbC\,\Ow$, and take nonzero $\xi\in\ker\Hw$ such that $\bktw{\xi,\Ow}=0$.
Since $\piw(\Aloc)\Ow$ is a core for $\Hw$ (see Theorem~6.2.4 of \cite{BR2}, Definition~3.1.17 and Corollary~3.1.20 of \cite{BR1}), there is a sequence $\hA_n\in\Aloc$ with $n=1,2,\ldots$ such that $\xi=\lim_{n\up\infty}\piw(\hA_n)\,\Ow$ and $\Hw\,\xi=\lim_{n\up\infty}\Hw\,\piw(\hA_n)\,\Ow$.
By substituting $\hA_n$ for $\hA$ in \rlb{A3} and letting $n\up\infty$ one gets
\eq
\bktw{Q_\Omega\,\xi,Q_\Omega\,\Hw\,\xi}\ge\gamma\,\snorm{Q_\Omega\,\xi}^2.
\en
Recalling that $\Hw\,\xi=0$ and $Q_\Omega\,\xi=\xi$, we find $0\ge\snorm{\xi}^2$ and hence $\xi=0$, which is a contradiction.

It readily follows from the assumption \rlb{gamma} that $\Hw$ has a gap above the ground state energy 0, or, more precisely, there is no spectrum of $\Hw$ in the interval $(0,\gamma)$.

It remains to prove that $\omega$ is pure.
Assume that $\omega$ is not pure.
Then there exists a state $\varphi$ that is distinct from $\omega$ and a constant $\lambda\in(0,1)$ such that $\lambda\varphi\le\omega$.
Since $\lambda\varphi$ is majorized by $\omega$, Theorem~2.3.19 of \cite{BR1} implies that there exists a nonnegative operator $T\in\piw(\OA)'$ with $\snorm{T}\le1$ such that
\eq
\lambda\,\varphi(\hA)=\bktw{\,T\,\Ow,\piw(\hA)\,\Ow},
\en
for any $\hA\in\OA$.
We now claim that $T\,\Ow\in\bbC\,\Ow$.
This implies $\varphi=\omega$, which is a contradiction.
To verify the claim, we recall that it is shown in Theorem~5.3.19 of \cite{BR2} that $e^{it\Hw}\in\piw(\OA)''$ for any $t\in\bbR$.
We then find
\eq
e^{it\Hw}\,T\,\Ow=T\,e^{it\Hw}\,\Ow=T\,\Ow,
\en
which implies $T\,\Ow\in\ker\Hw=\bbC\,\Ow$.~\qedm

\bigskip
{\small

It is my pleasure to thank Elliott Lieb for valuable discussions over many years, fruitful collaborations on quantum spin systems, and, most of all, his profound contributions to science, which have helped form the foundation of the modern mathematical physics of many-body systems.
I also thank Yoshiko Ogata for indispensable discussions and for allowing me to include her proof of Theorem~\ref{t:LUGGS} into the present article, 
Yuji Tachikawa
and
Haruki Watanabe for useful discussions and comments on the manuscript,
and 
Sven Bachmann,
Wojciech De Roeck,
Martin Fraas, 
Yohei Fuji,
Hosho Katsura,
Tohru Koma,
Taku Matsui,
Bruno Nachtergaele,
Masaki Oshikawa,
Ken Shiozaki,
and
Naoto Shiraishi
for useful discussions on related subjects.
The present work was supported in part by JSPS Grants-in-Aid for Scientific Research no. 22K03474.
}

%\newpage

\end{document}